\RequirePackage{etoolbox}
\patchcmd{\bibliographystyle}{#1}{abbrvnat}{}{}

\documentclass[a4paper,11pt]{article}
\pdfoutput=1 
\usepackage{jheppub} 
\usepackage[T1]{fontenc} 

\title{On the Inversion of High Energy Proton}

\author{Mikael Mieskolainen}
\affiliation{Department of Physics, Division of Particle Physics and Astrophysics, University of Helsinki, P.O. Box 64, 00014 Helsinki, Finland}

\emailAdd{mikael.mieskolainen@cern.ch}

\abstract{Inversion of the $K$-fold stochastic autoconvolution integral equation is an elementary nonlinear problem, yet there are no de facto methods to solve it with finite statistics. To fix this problem, we introduce a novel inverse algorithm based on a combination of minimization of relative entropy, the Fast Fourier Transform and a recursive version of Efron's bootstrap. This gives us power to obtain new perspectives on non-perturbative high energy QCD, such as probing the ab initio principles underlying the approximately negative binomial distributions of observed charged particle final state multiplicities, related to multiparton interactions, the fluctuating structure and profile of proton and diffraction. As a proof-of-concept, we apply the algorithm to ALICE proton-proton charged particle multiplicity measurements done at different center-of-mass energies and fiducial pseudorapidity intervals at the LHC, available on HEPData. A strong double peak structure emerges from the inversion, barely visible without it.}

\usepackage{amsmath}     
\usepackage{mathtools}   
\usepackage{algorithm}   
\usepackage{algorithmic} 
\usepackage{graphicx}
\usepackage{bm}
\usepackage{float}       
\usepackage{subcaption}  
\usepackage{feynmp}      
\graphicspath{{./paperfigs/}{./lhcfigs/}{./lhcfigs2D/}{./figs/}}

\def\multiset#1#2{\ensuremath{\left(\kern-.3em\left(\genfrac{}{}{0pt}{}{#1}{#2}\right)\kern-.3em\right)}}

\def\stirling#1#2{\ensuremath{\left\{\kern-.3em\genfrac{}{}{0pt}{}{#1}{#2}\kern-.3em\right\}}}

\usepackage{natbib}
\begin{document}
\noindent 
\maketitle

\newpage

\section{Introduction}

Differential distributions of final state particles in high energy collisions, such as multiplicity distributions, are notoriously difficult to model precisely from the first principles of QCD, due to the nature of confinement and non-perturbative infrared physics. The topic has been studied since the early days of Hagedorn fireball \cite{hagedorn1965statistical}, Feynman scaling \cite{feynman1969very} and Kobe-Nielsen-Olesen/Polyakov self-similarity (fractal) scaling \cite{polyakov1970hypothesis, koba1972scaling}. Hadron production has been treated resulting from break down of classic strings \cite{artru1974string}, Feynman-Field model \cite{field1978parametrization} and Lund strings of PYTHIA \cite{andersson1979semiclassical, andersson1983parton}, cluster hadronization of HERWIG \cite{webber1984qcd}, the topological expansion of cylindrical Pomeron particle production model or `quark-gluon strings' \cite{kaidalov1982quark}, to name some well known concepts. From the perturbative side, the local QCD parton-hadron duality \cite{azimov1985similarity} is perhaps to most appealing, with much support already from LEP data. Basically, it states that the hadronization happens at low virtuality scale independent of the hard scattering scale. However, then case of pure soft hadron-hadron scatterings without any hard scale involved is very interesting for collectivity or global color coherence reasons. These concepts are in a way or another implemented algorithmically in the Monte Carlo event generators, with varying number of free parameters to be fixed by data, with goals of factorizing universal, initial state and center-of-mass energy (in)-dependent properties.

In proton-nucleus $(pA)$ and nucleus-nucleus $(AA)$ physics a whole new class of phenomenological topics appear, from the space-time evolution of relativistic hydrodynamics and medium expansion to quark-gluon plasma signatures and the separation of different stages and collective phases of the process. A very interesting topic is the question which of these are due to heavy ion physics and which of them already happen in the elementary proton-proton interactions. In order to probe the ab initio physics behind many phenomenological models, and to factorize effects between heavy ions and elementary $pp$, solid mathematical approaches are crucial as analysis tools. 

In this work we introduce a novel inverse algorithm to invert a certain simple stochastic integral equation, the $K$-fold autoconvolution. By autoconvolution we mean convolving the distribution with itself and $K$-fold means repeating the operation recursively $K$-times. This is a nonlinear operation. In addition, we take the $K$ to be a random number. The `direct problem' and closely related problems are well studied in probability theory and implicitly ubiquitous in high energy physics, given the much studied pQCD logarithmic evolution integro-differential resummation schemes such as DGLAP in $\ln(Q^2)$ and BFKL in $\ln(1/x)$, which require as input the non-perturbative parton densities. The `inverse problem' has been studied much less, considering the variety of methods for standard deconvolution and more generally, methods to invert approximately linear and nonlinear integral equations with varying kernels. In experimental high energy physics, the unfolding of detector responses is more well known problem \cite{zech2016analysis, kuusela2015statistical,blobel2002unfolding,d1995multidimensional}. It corresponds to the usual deconvolution in the special case where the detector response matrix is of a convolution kernel type. One reason why inverse methods have not been studied extensively is, perhaps, the Monte Carlo event generators, which are extensively used as the direct models. The demanding  inverse problem in that case is the multidimensional parameter estimation or tuning of the event generator itself, instead.

The problem of inverting the pileup effects in inclusive soft QCD multiplicity measurements by inverting a $K$-fold autoconvolution was treated in \cite{matthews2011thesis}. However, the subtractive iterative solution there was proposed without a statistical treatment, explicit regularization or uncertainty estimation and only for a small number of $K$. We address these issues here by developing an all-order inverse. As far as we know, our approach which combines statistical modeling, Fast Fourier Transform (FFT) and Efron's bootstrap, is the first of its kind in high energy physics, but also including other fields. However, we mention that related problems have been studied also in a different context in laser optics \cite{gerth2014regularization}, inverse problems and mathematics \cite{gorenflo1994autoconvolution, choi2005iterative,meister2007optimal, van2007kernel, martin2009supremum}, and the problem is also closely related to the inverse Born series problem \cite{moskow2009numerical}.

In Section \ref{sec:directproblem} we introduce the mathematical direct problem and related formalism together with some simple examples and in Section \ref{sec:inverseproblem} and \ref{sec:algorithm} we go through the corresponding inverse problem, its discretization, regularization and the algorithmic solution. Section \ref{sec:simulations} is devoted to simulations and finally in Section \ref{sec:LHCdata} we do a proof-of-concept study of LHC-ALICE proton-proton multiplicity data. In Section \ref{sec:conclusions} we discuss further applications and research directions.

\section{Direct problem}
\label{sec:directproblem}

Let the underlying continuous or discrete differential distribution be $f_X$, with normalization $\int f_X(x)\,dx = 1$ and the corresponding random variable be $X$. The distribution can be discrete, for example, in a case of charged particle multiplicity spectrum measurements. The $K$-fold autoconvolution means that the measurement $g_Y$ is a sum of mutually independent identically distributed (i.i.d) random variables $Y = X_1 + X_2 + ... + X_K$. If we take $K \sim \text{Poisson}(\mu)$ and all $X_i \sim f_X$, this is known as a \textit{compound Poisson} distribution. The convolution arises here naturally, because the sum of random variables is equivalent to a convolution between the probability distributions of the corresponding random variables.

The autoconvolution operation is defined with
\begin{equation}
[f_X \circledast f_X] (y) \equiv \int_{-\infty}^\infty f_X(y-x) f_X(x)\,dx.
\end{equation}
If we take the support of $f_X$ on $[0,\infty)$, because physical observables are usually non-negative, then the autoconvolution integral range lower bound is zero. As a remark, in a more general setup, a translation dependent Green's function $G(x,y)$ would be used instead of $f(x-y)$. That would give a Fredholm's integral equation instead of a convolution integral, leading us to the `inverse Schr\"odinger equation' type of problems, as they are often called in the field of inverse problems.

For now, let us assume that the measurement follows a compound Poisson. However, in our algorithmic formulation we allow also any discrete distribution for $K$, not just Poisson distribution. The distribution of random variable $Y$ is now given, by construction, as an autoconvolution series weighted with Poisson probabilities
\begin{align}
\nonumber
g_Y(y) &= P_1f_X(y) + P_2 [f_X \circledast f_X](y) + P_3 [[f_X \circledast f_X] \circledast f_X](y) + \dots \\
\label{eq:directseries}
&= \frac{1}{1-e^{-\mu}}\sum_{n = 1}^\infty \frac{\mu^n }{n!} e^{-\mu} f_X^{\circledast^n} (y),
\end{align}
where the \textit{convolution power} $\circledast^n$ is defined recursively as $f^{\circledast^n} = f^{\circledast^{(n-1)}} \circledast f$ and $f^{\circledast 1}=f$. We have removed the unobservable case $n = 0$ which gives $Y = 0$ and renormalized the remaining Poisson probabilities $P_n, n = 1,2,3,\dots$ to sum to one. The zero suppressed mean of the variable $K$ is
\begin{equation}
\label{eq:muaverage}
\mathbb{E}[K > 0] = \mu/(1-\exp(-\mu)) > 1,
\end{equation}
where as the mean and variance (second central moment) of the compound distribution are given by useful formulas
\begin{align}
\label{eq:compoundmean}
\mathbb{E}[Y] &= \mathbb{E}[K]\mathbb{E}[X] \\
\label{eq:compoundvar}
\text{Var}[Y] &= \mathbb{E}[K]\text{Var}[X]+\text{Var}[K](\mathbb{E}[X])^2.
\end{align}
See \cite{grubbstrom2006moments} for all higher order moments and central moments of generic compound distributions. 

In the Fourier spectral domain, the characteristic function (CHF) $\varphi_X$ is defined as
\begin{equation}
\varphi_X(t) = \mathbb{E}[e^{itX}]=\int_\mathbb{R} e^{itX}\,dF_X(x) = \int_\mathbb{R} e^{itX}f_X(x)\,dx=\int_0^1e^{itQ_X(p)}dp,
\end{equation}
where $F_X(x)$ and $Q_X(p)$ are the cumulative and the inverse cumulative distribution functions of the random variable $X \in \mathbb{R}$. A probability generating function (PGF) $G_X$ corresponding to a discrete probability mass function $p \sim X \in \mathbb{N}$ is a Laurent series around zero defined as
\begin{equation}
G_X(z) = \mathbb{E}[z^X]=\sum_{n = -\infty}^\infty p(n)z^n.
\end{equation}
For the rest of the work we assume that the characteristic and generating functions are defined in the complex plane within their domain of convergence of the corresponding integral and sum, respectively, and leave the extensive algorithmic treatment of possible singularities for future work.

Using these tools, the well-known characteristic function for the compound Poisson is obtained with
\begin{align}
\nonumber
\varphi_Y(t) &= \mathbb{E}[e^{itY}] = \mathbb{E}_K \left[\left(\mathbb{E}[e^{it X}] \right)^K \right] = \mathbb{E}_K\left[ \left(\varphi_X(t)\right)^K \right] \\
\nonumber
&= \sum_{n=0}^\infty \varphi_X(t)^n P(K = n) \\
\nonumber
&= \sum_{n=0}^\infty \varphi_X(t)^n \frac{\mu^n}{n!}e^{-\mu} \\
&= e^{\mu(\varphi_X(t) - 1)}
\end{align}
and solving similarly for the case $K > 0$ ($n=1,2,...$) gives
\begin{align}
\label{eq:nonlinearmap}
\varphi_{g}(t) \equiv \varphi_{Y|K>0}(t) = \frac{e^{-\mu}\left( e^{\mu \varphi_f(t)}-1\right)}{1-e^{-\mu}} = \frac{1}{e^\mu - 1}(e^{\mu \varphi_f(t)} - 1),
\end{align}
where we denoted the characteristic function of $f_X(t)$ with $\varphi_f(t)$. A direct attempt to invert this would be to take the inverse Fourier transform, but we would encounter the problem of multivalued complex logarithm. Defining the complex logarithm in a suitable way was proposed together with a kernel estimator, for `decompounding' in \cite{van2007kernel}. Here we take a different approach, where we \textit{avoid} altogether using complex logarithms or other multivalued complex operations such as complex roots. 

We will obtain a practical formal notation for the algorithm by representing the problem with an operator $\mathcal{F} : \Omega_X \rightarrow \Omega_Y$, where $\Omega_X$ and $\Omega_Y$ are the original and the smeared domain
\begin{equation}
\label{eq:continousdirectmap}
\mathcal{F}(f) + \delta g = g.
\end{equation}
Our main goal is to invert this nonlinear mapping taking also into account the statistical fluctuations $\delta g$. The nonlinearity comes directly from the fact that the operator $\mathcal{F}$ depends on $f$.

To point out, the existence of $K$-fold convolution is guaranteed by the \textit{infinitely divisibility} of a probability distribution, which is the L\'{e}vy-Khintchine theorem. That is, for given a distribution $f$, there exist for any $n$, the corresponding $n$-th convolution power. All members of the stable family - for example Gaussian, Poisson, Negative binomial, Gamma, Cauchy (non-relativistic Breit-Wigner), L\'{e}vy and Landau - are infinite divisible. They are also closed under convolution.

\subsection*{Examples}

An interesting phenomenological fact, already observed in the CERN proton-antiproton UA5 data at $\sqrt{s} = 540$ GeV \cite{giovannini1986negative, alner1985multiplicity}, is that the charged particle multiplicity distributions tend to follow approximately negative binomial distribution (NBD) or two of them, as the classic two component models may suggest. Same observation is still approximately valid at the LHC \cite{ghosh2012negative}. Mathematically speaking NBD follows from a compound Poisson distribution of number of $K$ \textit{logarithmic series} $f(n;p) = p^n/(-n\ln(1-p)), n \in \mathbb{N}_+$ i.i.d variables each with shape parameter $p \in (0,1)$ while $K$ being Poisson distributed with parameter $\mu \in \mathbb{R}_+$. Thus we obtain negative binomial distribution on $\mathbb{N}_+$ as the compound distribution with parameters $1-p \in (0,1)$ and $-\mu/\ln(1-p) \in \mathbb{R}_+$. This mathematical picture may be useful to keep in mind while thinking the possible dynamical explanations.

Monte Carlo models reproduce NBD like distributions by multipomeron cuts or multiparton interaction modeling -- the low multiplicity shape of the distribution is usually \textit{very sensitive} to the non-perturbative proton profile or eikonal density driving the number $K$ with impact parameter dependent distributions, and also to diffraction contribution. Kinematically, a system with the cms energy $W$ will span a longitudinal rapidity range $Y \sim \ln(W^2/W_0^2)$, where $W_0$ is order of proton mass. In Lund model like scenarios, also the average number of particles scales like $\langle N \rangle \sim \ln(W^2/W_0^2)$. A multistring system total multiplicity consisting of $K$-strings will behave as $\langle N \rangle \sim \sum_{i=0}^{K-1} \ln(s_{i,i+1}^2/W_0^2)$ with $s_{i,i+1} = (k_i + k_{i+1})^2$ being sub-Mandelstam invariants constructed from the parton 4-momenta $\{k_i\}$ spanning the strings and the variance Var$[N]$ behaves proportionally to $\langle N \rangle$ \cite{andersson1989local}, which is Poisson like behavior. So if the sub-$W^2$ distributions behave typically like falling powerlaws and the number of $K$-simultaneous strings is also modeled as Poisson like with impact parameter dependence or not, one will generate negative binomial like distributions. To point out at this point, the soft transverse momentum dependence with multiplicity is not understood from the first principles.

\begin{figure}[h]
\begin{center}
\includegraphics[width=90mm]{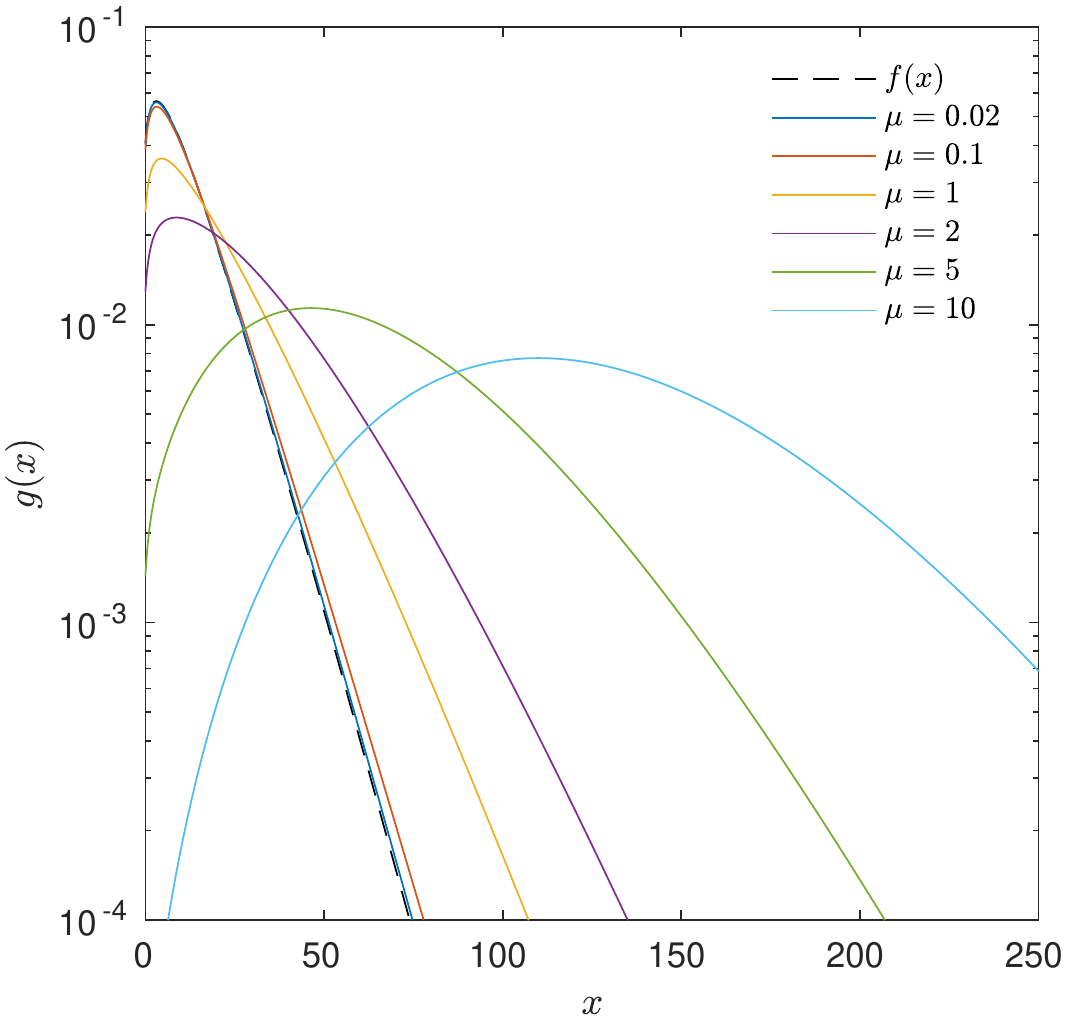}
\end{center}
\caption{$K$-fold compound Poisson autoconvolution of the negative binomial distribution with varying $\mu$. }
\label{fig:analyticNBD}
\end{figure}

Now let us think instead the experimentally visible superposition of `pileup' proton-proton interactions at the LHC for a given Poisson $\mu$. At the LHC Poisson $\mu$ is measured for low values of $\mu \sim 0.05$ (ALICE) using the minimum bias trigger occupancy\footnote{$\langle R \rangle_{t,N_B} = \frac{\frac{1}{\Delta t} \int_{\Delta t} f_R(t) \,dt}{N_B f_O}$, where $N_B$ is the number of colliding bunch pairs, $f_O$ the LHC revolution frequency (Hz) and $f_R(t)$ the instantaneous trigger rate (Hz).} $\langle R \rangle = 1 - P_0(\mu) = 1 - e^{-\mu} \Leftrightarrow \mu = -\ln(1 - \langle R \rangle)$, where $R \in [0,1]$. For large values of $\mu \sim 50$ when the occupancy is completely saturated (ATLAS, CMS), the scaling of the track multiplicity or number of primary vertices is used to determine $\mu$. A combination of both can be useful for intermediate values of $\mu \sim 5$ (LHCb), for example. The pileup is illustrated in Figure \ref{fig:analyticNBD}. Convolution of a negative binomial with a negative binomial is yet again a negative binomial -- this is the closure under convolution. It is worth pointing out that the estimation of $\mu$ based solely on the measured distribution $g(x)$ itself cannot be done in general. This is because if the distribution is from the family of distributions which are closed under convolutions, then we have no capability of identifying if the measured distribution is pileup free, or, one with a larger scale and position parameter than the expected one, or the autoconvoluted version of one with smaller scale and position parameter. Thus, $\mu$ must be a parameter of the inversion and must be inferred from other measurements or theoretical constructions.

Clearly, our discussion is now very close to the KNO / Polyakov scaling \cite{koba1972scaling}, which states that there is one elegant scaling function $\psi$ giving the probability of multiplicity $n$ as a function of the scaling variable $n/\langle n(s) \rangle$
\begin{equation}
P(n;s) = \frac{1}{\langle n(s) \rangle} \psi \left(\frac{n}{\langle n(s) \rangle} \right),
\end{equation}
where $s$ is the the center of mass energy squared Mandelstam variable, a Lorentz scalar. However, this scaling is known to be violated at some level since the ISR times, and at the LHC it holds merely in limited regions of the full phase space, in a narrow central rapidity window. In general, when inspecting scaling or its violation one needs to take into account the collision initial state, e.g. $pp$ versus $e^+e^-$, the fiducial phase space of the measurement and contribution of diffraction. At model level, multiparton or multipomeron exchanges running as a function of energy are usually creating significant deviations from the KNO scaling. Clearly additional effects arise from soft versus hard scale  interactions. See the books \cite{kittel2005soft,dokshitzer1991basics} for discussion on a variety of perturbative QCD and other scenarios exhibiting scaling.

By central limit theorem, the $K$-fold convolution will map almost\footnote{For example Cauchy / non-relativistic Breit-Wigner type distributions do not turn into Gaussian, also moments for these distributions are undefined.} every distribution eventually into a Gaussian distribution when $\mathbb{E}[K] \rightarrow \infty$.  This is clearly visible in Figure \ref{fig:analyticNBD}, where we used the direct mapping of Equation \ref{eq:directseries} with different Poisson $\mu$-values to convolve the negative binomial distribution
\begin{equation}
P_{\text{NBD}}(n \,;\, \langle n \rangle,k) = \frac{\Gamma(k+n)}{\Gamma(k)\Gamma(n+1)}\left[\frac{\langle n \rangle}{k+\langle n \rangle}\right]^n \left[\frac{k}{k+\langle n \rangle}\right]^k
\end{equation}
with a fit to the LHC charged particle multiplicity data in proton-proton at $\sqrt{s} = 7$ TeV for pseudorapidity range $\eta \in [-0.9, 0.9]$ obtained from \cite{ghosh2012negative}. The fit parameters central values are $\langle n \rangle = 12.5$ and $ k = 1.4$ , where $\langle n \rangle$ is the average multiplicity of NBD and $k$ is related to variance $D^2 = \langle n^2\rangle - \langle n \rangle$ by $D^2/\langle n \rangle^2 = 1/\langle n \rangle + 1/k$.

In an explicit experimental realization, in order to take into account the detector response and its effects on the resolution and efficiency, an unfolding algorithm with uncertainty estimation and other corrections should be processed first -- if necessary. Output of this can be then propagated to the algorithm described here. For the rest of the work, we assume this to be the case.

\section{Inverse problem}
\label{sec:inverseproblem}

Solving the $K$-fold autodeconvolution means here a statistical inverse of Equation \ref{eq:continousdirectmap}. Unfortunately the measurement noise and modeling errors will usually always forbid us executing a direct inversion in practice, even if we would have a closed form expression to do so. This ill-posedness requires usually either implicit or explicit regularization which turns the problem into a more well-posed one.

We calculate the convolution by pointwise multiplication in Fourier domain using the convolution theorem $f \circledast g = \mathfrak{F}[f]\mathfrak{F}[g]$, which works also the other way around $\mathfrak{F}[f]\circledast \mathfrak{F}[g] = f g$, which is also often called ambiguously convolution or folding of distributions of $f$ and $g$. Now, a fixed $K \in \mathbb{N}$ autoconvolution is given by
\begin{align}
\mathfrak{F}[f^{\circledast K}] &= \left( \mathfrak{F}[f] \right)^K = \mathfrak{F}[g],
\end{align}
where power is applied as $z^K = (|z|e^{i(\varphi + 2\pi n)})^K = |z|^Ke^{iK\varphi}e^{2\pi i n K}, n \in \mathbb{Z}$. Thus, in spectral domain the stochastic autoconvolution results in random scaling of amplitudes and rotation of phases of the characteristic function. Now when $K$ is not an integer this results in a multi-valued operation, because $e^{2\pi i n K}$ does not map a full rotation around the complex plane for different values of $n$. To remind, if we shift the probability density $f(x-x_0)$, this results in a phase shift in the frequency domain by $\exp(-i\omega x_0)\mathfrak{F}[f]$.

A formal `naive' inverse is obtained by taking the complex $K$-th root in Fourier domain, and then proceeding with an inverse Fourier transform
\begin{align}
\left( \mathfrak{F}[f]^K \right) ^{1/K} &= \left( \mathfrak{F}[g] \right)^{1/K} \\
f &= \mathfrak{F}^{-1}\left[ \left( \mathfrak{F}[g] \right)^{1/K} \right],
\end{align}
analogously to the standard naive Fourier deconvolution. However, complex root is a multivalued operation $|z|^{1/K}e^{i(\text{Arg}(z) + 2\pi n)/K}$ with $n = 0,1,\dots,K-1$ roots with suitable branch cuts to be specified algorithmically, in principle. Now we see that the only strictly well posed case is the case when the characteristic function is real and non-negative $\leftrightarrow$ a mirror symmetric (even) probability density around zero by the properties of Fourier transform, a case which is clearly too restrictive to be of interest here. Also, this `solution' does not take into account that $K$ is a random variable, neither it takes into account the finite statistics of $g$ and mismatches between the measurement and modeling causing instabilities to be regulated. For some related  discussion, see for example \cite{bogsted2010decompounding}.

\subsection*{Discretization}
\label{sec:discretization}

In order to solve the finite sample version of the problem, we need a (discrete) representation of the problem. For simplicity, computational efficiency and due to the high energy physics analysis convention -- as a practical format of the measurement, we use histograms as the density estimators of $f$ and $g$, even if simple kernel (Parzen) estimators and even more advanced density estimation techniques could be used in principle. Also, we consider only 1D-version of the problem. However, extension to higher dimensions is formally straightforward.

The histograms are defined in terms of $D, KD$ bins with corresponding fixed bin width $\Delta x \equiv \Delta y$. In the context of algorithms histograms are represented with finite non-negative column vectors, which we denote with bolded symbols
\begin{equation}
\mathbf{f}[n] = \left| \{f_i \in [x_n,x_n+\Delta x) \} \right|, \;\; n \in \mathbb{Z}_{\Omega_X} := \{0,1,\dots,D-1\}
\end{equation}
and similarly for a $K$-fold autoconvoluted
\begin{equation}
\mathbf{g}[n] = \left| \{g_i \in [y_n,y_n+\Delta y) \} \right|, \;\; n \in \mathbb{Z}_{\Omega_Y} := \{0,1,\dots,K(D-1) \},
\end{equation}
where $f_i,g_i \in \mathbb{R}$ denote sample elements and $x_n,y_n$ denote the bin lower edges. We extended the domain of autoconvoluted by $K$-times and because $K$ is a random variable, in practice we take some large enough integer which avoids any truncation problems in the upper tail. Only values of $\{g_i\}$ are directly observable and measured. Now the discrete autoconvolution is defined as
\begin{equation}
[\mathbf{f} \circledast \mathbf{f}] [n] = \sum_{m = 0}^{2(D-1)} \mathbf{f}[m]\mathbf{f}[n-m],
\end{equation}
and the $K$-fold version follows recursively. In the algorithmic implementations, when necessary to extend the domain of vectors due to convolution, we do it explicitly by zero padding.


In principle, the binning can be also non-uniform, such as logarithmic. However, that must be taken explicitly into account in the evaluation of the discrete convolution, which is defined using uniform sampling. One classic solution could be interpolation and resampling schemes with B-splines, for example. Here, however, we use only fixed binning. The bin width $\Delta x$ needs to be chosen such that the spectrum resolution criteria and event count statistics are taken into account. This in order not the create `noisy spectrum' with low bin counts, or on the other hand, induce a loss of resolution. Thus, it has also a role as an implicit regulator.

\subsection*{Relative entropy minimization}
\label{sec:relative-entropy}

The measured histogram $\mathbf{g}$ is assumed to be bin-by-bin Poisson distributed, a usual assumption, given also the Poisson superposition property $\sum_i \text{Poisson}(\mu_i) \sim \text{Poisson}\left(\sum_i \mu_i \right)$. The Poisson likelihood with a known constant background histogram $\mathbf{b}$ is then written as a product over the histogram bins
\begin{equation}
p(\mathbf{g}|\mathbf{f}) = \prod_i \frac{[\mathcal{F}\mathbf{f} + \mathbf{b}]_i^{g_i} e^{-[\mathcal{F}\mathbf{f} + \mathbf{b}]_i}}{g_i!}
\end{equation}
and its negative log-likelihood is now our fit quality term
\begin{align}
J(\mathbf{f}|\mathbf{g}) = -\log p(\mathbf{g}|\mathbf{f}) &= \sum_i -g_i \log([\mathcal{F}\mathbf{f} + \mathbf{b}]_i) + [\mathcal{F}\mathbf{f}]_i + b_i + \log(g_i!)
\end{align}
and by neglecting terms which do not affect the solution and rearranging gives
\begin{align}
J(\mathbf{f}|\mathbf{g}) = \sum_i g_i \frac{g_i}{\log([\mathcal{F}\mathbf{f} + \mathbf{b}]_i)} + [\mathcal{F}\mathbf{f}]_i - g_i.
\end{align}
This often used formulation is the minimum entropy solution, and is equivalent to minimizing generalized Kullback-Leibler divergence (relative entropy) \cite{kullback1951information}, known also as the Csisz\'{a}r I-divergence \cite{csiszar1991least}. The gradient of this functional is
\begin{equation}
\label{eq:RLgradient}
\nabla J(\mathbf{f}) = \frac{\partial}{\partial \mathbf{f}}\left(-\mathbf{g}^T\log(\mathcal{F}\mathbf{f} + \mathbf{b}) + 1^T \mathcal{F}\mathbf{f} \right) = \mathcal{F}^T\mathbf{1} - \mathcal{F}^T \frac{\mathbf{g}}{\mathcal{F}\mathbf{f} + \mathbf{b}} = \mathcal{F}^T \left( \mathbf{1} - \frac{\mathbf{g}}{\mathcal{F}\mathbf{f} + \mathbf{b}} \right),
\end{equation}
where divisions are understood component wise. If distributions are normalized to unity, then $\mathcal{F}^T\mathbf{1} \equiv \mathbf{1}$ holds. Here, we use event counts.

The gradient based iterative update or a fixed point iteration is now obtained from the steepest descent update $\mathbf{f}_{k+1} = \mathbf{f}_k - \gamma D_k\nabla J(\mathbf{f}_k)$, where $D_k$ is a gradient scaling matrix. If we set $D_{\mathbf{f}_k} = \text{diag}(\mathbf{f}_k)$ and $\gamma = 1$ which is well known to correspond in this case to the expectation maximization (EM) \cite{dempster1977maximum} formulation of a frequentist maximum likelihood solution under the Poisson likelihood, then we obtain a Richardson-Lucy \cite{richardson1972bayesian,lucy1974iterative} type multiplicative deconvolution formula well known in optics and astrophysics
\begin{equation}
\mathbf{u}_k = \left( \mathbf{f}_k \oslash (\mathcal{F}^T \mathbf{1}) \right)\odot \left( \mathcal{F}^T \mathbf{g} \oslash (\mathcal{F} \mathbf{f}_k + \mathbf{b} \right),
\end{equation}
where $\odot$ and $\oslash$ are Hadamard's vector component wise product and division, respectively. This was also re-invented by D'Agostini \cite{d1995multidimensional} in high energy physics unfolding context. For the rest of the paper, we set the background $\mathbf{b} = 0$. The RL-type update conserves non-negativity of the solution and the total number of events. Another way to derive the RL type update is to set gradient Equation \ref{eq:RLgradient} to zero, move terms on both sides, multiply by $\mathbf{f}$ and do fixed point iteration.

\subsection*{Regularization}
\label{sec:regularization}

The problem of choosing the spectral bandwidth cut-off or regularization strength in the non-transformed domain are the most common but also the most difficult topics of inverse problems. Motivated by the fact that the distributions of observables for us are usually smooth and non-discontinuous, we use a traditional variational regularization with a Tikhonov $\ell_2$-norm type regularity functional
\begin{equation}
J_R(f) = \lambda \int_\Omega \|\nabla f(x)\|^2 \, dx,
\end{equation}
where $\Omega \subset \mathbb{R}$. The minimum of this is given by
\begin{align}
\nonumber
\nabla J_R(f) &= \lambda \nabla \left(\int_\Omega \|\nabla f(x)\|^2 dx \right) \\
&= \lambda \nabla^* \nabla f(x) = -\lambda \text{div}(\nabla f(x)) = -\lambda\nabla^2 f(x) = 0,
\end{align}
where we applied to the integral von Neumann boundary conditions and Gauss divergence-theorem, the standard vector analysis identities and $\nabla^2$ is the usual Laplacian. That is, the gradient descent direction is given by the negative Laplacian.

Instead of using the abstract functional gradient, the regularization term is discretized using a finite difference matrix and the corresponding finite Laplacian
\begin{equation}
\nabla_M \equiv 
\begin{pmatrix}
1 & -1 & 0 & \cdots & 0 \\
0 &  1 & -1 & & \vdots \\
\vdots &  & & \ddots & 0 \\
0 & \cdots & 0 & 1 & -1
\end{pmatrix}, \; 
\nabla_M^2 \equiv \nabla_M^T \nabla_M =
\begin{pmatrix}
-1 & 2 & -1 & \cdots & 0 \\
0 &  -1 & 2 & -1 & \vdots \\
\vdots &  & & \ddots & 0 \\
0 & \cdots & -1 & 2 & -1
\end{pmatrix}.
\end{equation}
It is naturally also possible to modify the boundary conditions which affect these matrices, based on a priori assumed or known properties of the distribution $f(x)$.

Regularization parameter $\lambda$ depends on the number of observed event count $N$, the number of histogram bins $D$ (= problem discretization) and the shape of the distribution of interest. The average pileup $\mu$ and the functional smoothness of the distribution under inversion are very important factors affecting the nonlinear direct operator and thus the necessary regularization strength. In the limit $N_E \rightarrow \infty$ and $\mu \rightarrow 0$, we can take asymptotically $\lambda \rightarrow 0$. There are several semi-heuristic `data-driven' methods developed for selecting the regularization. The well known principles are the L-curve, generalized cross validation (GCV), Morozov's discrepancy principle and various covariance and residual spectrum analysis methods \cite{mueller2012linear}. Monte Carlo event generator driven or a toy model based analysis together with data driven approaches is also a reasonable choice in high energy physics. We use a simple variational equilibrium approach to choose the regularization parameter. This is is described in Section \ref{sec:simulations}.

The full solution taking together fidelity + regularity usually means directly minimizing in the direction of combined gradient of the full cost
\begin{equation}
\min_\mathbf{f} \,\, J(\mathbf{f}|\mathbf{g}) + \lambda J_R(\mathbf{f})
\end{equation}
as implemented for example in \cite{bardsley2004nonnegatively} in the context of image deconvolution. However, this we observed to give very unstable results in this problem. Thus, we implemented the regularization step through forward-backward slitting type gradient update, similar in fashion to proximal optimization algorithms. It means that we update the result after the EM-update step as
\begin{equation}
\mathbf{f}_{k+1} = \mathcal{P}_+\left(\mathbf{u}_k - \lambda\nabla_M^2\mathbf{u}_k \right),
\end{equation}
where $\mathcal{P}_+(\cdot)$ is the positive solution projector operator $\mathcal{P}_+ : \mathbb{R}^n \rightarrow \mathbb{R}_+^n$ such that $\mathcal{P}_+(\mathbf{x}) = \mathbf{y}$ gives $y_i = x_i$ if $x_i > 0$, else, $y_i = 0, \; \forall i = 1,\dots,n$. Separating the regularization as a separate step resulted in a much improved behavior. However, the approach chosen here is the best performing we found from the vast phase space of optimization algorithms and regularization approaches. One must remember that we are not dealing here with an inverse problem with a fixed kernel, thus this problem has its own peculiarities.

\newpage

\section{Algorithm}
\label{sec:algorithm}

For every iteration, we construct the autoconvolution operator by taking the discrete Fourier transform using FFT as $F = \mathfrak{F}[\mathbf{f}]$ and then construct the empirical characteristic function of the projected measurement with a nonlinear complex map $G = \frac{1}{e^\mu - 1}\left( e^{\mu F} - 1 \right)$ using Equation \ref{eq:nonlinearmap}, where complex exponential is the single valued function $\exp(z) = \sum_{n \geq 0}z^n/n!, \, z \in \mathbb{C}$. The autoconvolution `kernel' in spectral domain is then obtained by taking a point wise division $H = G \oslash F$, inverse transforming $\mathbf{h} = \mathfrak{F}^{-1}[H]$ and finally representing it as a Toeplitz convolution matrix to obtain $\mathcal{F} = \mathcal{T}[\mathbf{h}]$. The operator $\mathcal{T}$ represents here a simple standard way to construct a Toeplitz matrix. We assume $F$ to be non-zero for all of its components, which might be violated in pathological cases. These are intrinsically non-invertible without extra a priori information.

We remark here that this nonlinear map and division in spectral domain has a different goal than in the usual Fourier domain linear deconvolution such as in Wiener filters \cite{wiener1949extrapolation}, where the noise amplification is regularized in the spectral domain. In Wiener filtering, the division gives a solution to the problem non-recursively. Here, on the other hand, we use the spectral domain as a way to obtain the Toeplitz matrix representation in the probability domain extremely efficiently via FFT. The full solution is recursive and regularized in the probability domain.

\begin{algorithm}
\textbf{INPUT:} Smeared histogram $\mathbf{g}$ with $N$ events, regularization strength $\lambda \in \mathbb{R}_+$, Poisson mean $\mu \in \mathbb{R}_+$ \textbf{OR} discrete distribution $P_n$ with $\sum_{n=1}^{\tilde{K}} P_n = 1$, number of iterations $R \in \mathbb{N}_+$, background histogram $\mathbf{b}$ or otherwise $\mathbf{b} = 0$.\\
\begin{algorithmic}
\STATE \textbf{If} $\mu$ given, use option \textbf{\S I} below, \textbf{else}, use input $P_n$ and option \textbf{\S II} below
\vspace{0.25em}
\STATE Initialize $\mathbf{f}_1 \leftarrow \mathcal{Z}_{\tilde{K}}(\mathbf{g - b})$ with ${\tilde{K}}$-fold zero padding $\mathcal{Z}_K(\cdot)$
\vspace{0.5em}
\FORALL{$k = 1,\dots,R$}
\STATE \textbf{Step 1.} Construction of the direct operator: \\
\vspace{0.5em}
\STATE \textbf{\S I}: $\,$ Poisson case; `all-order' map: \\
$F \leftarrow \mathfrak{F}[\mathbf{f}_k]$,
$G \leftarrow \frac{1}{e^{\mu}-1}\left( e^{\mu F}-1 \right)$,
$H \leftarrow G \oslash F$,
$\mathbf{h} \leftarrow \mathfrak{F}^{-1}[H]$,
$\mathcal{F}_k \leftarrow \mathcal{T}[\mathbf{h}]$
\vspace{0.5em}
\STATE \textbf{\S II}: Generic $P_n$ case; evaluated `order-by-order': \\
\FORALL{$n = 2,\dots,{\tilde{K}}$}
\STATE $F_k^{(n)} \leftarrow \mathcal{T}[\mathbf{f}_k^{\circledast^{n-1}}]\,\,$ \# Construct convolution Toeplitz matrix
\STATE $\mathbf{f}_k^{\circledast^n} \leftarrow \mathbf{f}_k^{\circledast^{n-1}} \circledast \mathbf{f}_k$ \# Calculate convolution power
\ENDFOR
\STATE $\mathcal{F}_k \leftarrow \sum_{n=1}^{\tilde{K}} P_n F_k^{(n)}$, where $F_k^{(1)} = I$ \# Weighted sum of convolution matrices
\STATE
\STATE \textbf{Step 2.} EM-step + regularization/smoothing with positivity constraint:
\STATE $\mathbf{u}_{k} \leftarrow \left( \mathbf{f}_k \oslash (\mathcal{F}^T \mathbf{1}) \right)\odot \left( \mathcal{F}^T \mathbf{g} \oslash (\mathcal{F} \mathbf{f}_k + \mathbf{b} \right)$
\STATE $\mathbf{f}_{k+1} \leftarrow \mathcal{P}_+\left( \mathbf{u}_k - \lambda \nabla_M^2 \mathbf{u}_k \right)$
\ENDFOR
\STATE
\end{algorithmic}
\textbf{OUTPUT:} Deconvolution estimate $\hat{\mathbf{f}} \leftarrow \mathcal{Z}_{\tilde{K}}^{-1}(\mathbf{f}_R)$ with the same support as $\mathbf{g}$.
\caption{\textsc{Kisu}: K-fold Inverse of Stochastic Autoconvolution}
\label{algo:algorithmSAIA}
\end{algorithm}

Discretization of the operator $\mathcal{F}$ in the case of arbitrary (non-Poisson) compound distribution is done with a finite number of convolution Toeplitz matrices $F^{(n)}, n = 1,\dots,{\tilde{K}}$ with $F^{(1)} = I$ up to a numerically suitable finite order ${\tilde{K}}$. The maximum order is basically limited only by the available computing resources. These Toeplitz matrices are then added together with weights $P_n$. That is, we evaluate the discretized and truncated version of Equation \ref{eq:directseries}. This numerical finite order implementation was cross checked against the exact all order Fourier domain method in the Poisson case, as described above, and they agreed within floating point accuracy.

Now summarizing: the multiplicative algorithm describe above corresponds to a non-negative solution with Poisson bin fluctuations likelihood with the generalized Kullback-Leibler divergence being the fit criteria. On the other hand, subtractive algorithms are usually encountered with iterative solutions to unconstrained optimization under Gaussian noise and $\ell_2$-minimum norm minimizing the least squares cost $J_{\text{LS}}(\mathbf{f}) = \frac{1}{2} \|\mathcal{F}\mathbf{f} - \mathbf{g}\|^2$ with an iterative solution
\begin{equation}
\mathbf{f}_{k+1} = \mathbf{f}_k - \gamma \nabla J_{\text{LS}} = \mathbf{f}_k - \gamma \mathcal{F}^T \left( \mathcal{F} \mathbf{f}_k - \mathbf{g} \right)
\end{equation}
which is known as the Landweber iteration scheme \cite{landweber1951iteration}. Non-negativity of the solution should be enforced simply with the projector $\mathcal{P}_+(\cdot)$, because it is not a built-in constraint of the standard least squares. The $\gamma$ is a relaxation parameter which controls the convergence, and can be optimized by several different line search methods or by fixing it to a constant $\sim 1$, which gives (much) slower convergence. This Gaussian version of the algorithm could be utilized with large event samples. With low event count histograms, the multiplicative solution was observed to be superior in terms of stability of the solutions.

\subsection*{Uncertainty estimation}

Point estimates are obtained with Algorithm \ref{algo:algorithmSAIA}, which we call by the name \textsc{Kisu}. The algorithm first solves the direct problem and then one EM-step plus regularization of the inverse, and recursively alternate between these two until convergence. The regularization is done explicitly, because it is well known that the EM-type maximum likelihood solution alone will amplify the high frequencies ($\sim$ Poisson noise) when the number of iterations $k \rightarrow \infty$. That is, iteration first fits the low frequency Fourier components and later starts to fit the high frequency noise to the solution. Semi-explicit regularization strategy would correspond to an early iteration stop. However, with an explicit regularization scheme as here, the iteration can be allowed to continue till convergence within some numerical threshold.

\begin{algorithm}
\textbf{INPUT:} Smeared histogram $\mathbf{g}$, number of bias correction iterations $M \in \mathbb{N}_+$, bootstrap sample size per iteration $B \in \mathbb{N}_+$ and Algorithm \textbf{A\ref{algo:algorithmSAIA}} parameters.\\
\begin{algorithmic}

\STATE Run \textbf{A\ref{algo:algorithmSAIA}} using $\mathbf{g}$ as input, obtain $\hat{\mathbf{f}}_0$
\STATE Set $\hat{\mathbf{f}}_\star \leftarrow \hat{\mathbf{f}_0}$, $\hat{\mathbf{g}}_\star \leftarrow \mathbf{g}$ 
\FORALL{$i = 1,\dots,M$}
\STATE Set $S_i \leftarrow \emptyset$
\FORALL{$j = 1,\dots,B$}
\STATE Draw $\mathbf{g}_j^* \sim \hat{\mathbf{g}}_\star$ with replacement using toy Monte Carlo
\STATE Run \textbf{A\ref{algo:algorithmSAIA}} using $\mathbf{g}_j^*$ as input, obtain $\mathbf{f}_j^*$, update $S_i \leftarrow S_i \cup \{\mathbf{f}_j^*\}$ 
\ENDFOR
\STATE \textbf{Step 1.} Iterated bias estimate using bootstrap  sample $S_i$ mean:\\
$\hat{\text{Bias}}[\hat{\mathbf{f}}] \leftarrow \text{Mean}[S_i] - \hat{\mathbf{f}}_\star$
\STATE \textbf{Step 2.} Point estimate with bias correction, and non-negativity re-enforced:\\
$\hat{\mathbf{f}}_\star \leftarrow \mathcal{P}_+(\hat{\mathbf{f}}_0 - \hat{\text{Bias}}[\hat{\mathbf{f}}])$
\STATE \textbf{Step 3.} Using direct operator estimation \textbf{Step 1.} of \textbf{A\ref{algo:algorithmSAIA}}, generate:\\
$\hat{\mathbf{g}}_\star \leftarrow \mathcal{F}_\star(\hat{\mathbf{f}}_\star)$
\ENDFOR
\STATE
\end{algorithmic}
\textbf{OUTPUT:} Bias corrected estimate $\hat{\mathbf{f}}_\star$, first estimate $\hat{\mathbf{f}}_0$ and the sample $S_B$ for the standard bootstrap confidence interval evaluation.
\caption{`Daughter bootstrap' -- Iterative Bias Correction}
\label{algo:algorithmDAUGHTER}
\end{algorithm}

The iterative bias estimation and uncertainty estimation are done using numerical Monte Carlo bootstrap with Algorithms \ref{algo:algorithmDAUGHTER} and \ref{algo:algorithmMOTHER}.  We call these daughter and mother bootstrap, respectively. Algorithm is started by calling the mother bootstrap, which calls the daughter bootstrap, which calls \textsc{Kisu}. Bootstrap is selected here due to nonlinear and recursive nature of the problem which makes the usual linearized Taylor expansion error propagation and analytical (Gaussian) approximations unreliable or semi-impossible. Bootstrap was introduced in the seminal paper by Efron in 1979 \cite{efron1979bootstrap} and the recursive bias correction was first discussed in 1986 \cite{hall1986bootstrap}. In high energy physics unfolding context, bootstrap iterated bias correction and related confidence intervals were utilized only quite recently in \cite{kuusela2015statistical} and we follow a similar strategy. The underlying inverse problem is nonlinear here, in contrast. Bootstrap is relatively easy method to implement and variants of it have already long been used in high energy physics dubbed often under the large umbrella of `toy Monte Carlo'. A practical problem is the high computational cost, fortunately bootstrap sampling is trivially parallelizable. Although Algorithm \ref{algo:algorithmSAIA} and iterative bias correction loop of Algorithm \ref{algo:algorithmDAUGHTER} are recursive, and thus cannot be made fully parallel.

\begin{algorithm}
\textbf{INPUT:} Smeared histogram $\mathbf{g}$, bootstrap sample size $Q \in \mathbb{N}_+$, Algorithm \textbf{A\ref{algo:algorithmDAUGHTER}} and \textbf{A\ref{algo:algorithmSAIA}} parameters.\\
\begin{algorithmic}
\STATE Set $S_\star \leftarrow \emptyset, S_0 \leftarrow \emptyset$
\FORALL{$q = 1,\dots,Q$}
\STATE Draw $\mathbf{g}_q^* \sim \mathbf{g}$ with replacement using toy Monte Carlo
\STATE Run \textbf{A\ref{algo:algorithmSAIA}} using $\mathbf{g}_q^*$ as input, obtain $\mathbf{\hat{f}}_{\star q}^*$ and $\mathbf{\hat{f}}_{0 q}^*$, update $S_\star \leftarrow S_\star \cup \{\mathbf{\hat{f}}_{\star q}^*\}, S_0 \leftarrow S_0 \cup \{\mathbf{\hat{f}}_{0 q}^*\}$
\ENDFOR
\STATE
\end{algorithmic}
\textbf{OUTPUT:} Samples $S_\star$ and $S_0$ for calculating the confidence intervals, median etc., for both bias corrected $(\star)$ and uncorrected $(0)$ estimates.
\caption{`Mother bootstrap'}
\label{algo:algorithmMOTHER}
\end{algorithm}

Bias of a parameter estimator is by definition $\text{Bias}[\hat{\theta}] = \mathbb{E}[\hat{\theta}] - \theta$. Iterative bootstrap estimate of the bias replaces the unknown expectation with a bootstrap sample mean and the true parameter value with the current estimate of the parameter. This is described in algorithm \ref{algo:algorithmDAUGHTER}. However, the variance of the estimator $\text{var}[\hat{\theta}]=\mathbb{E}[(\hat{\theta})-E[\hat{\theta}])^2]$ can grow as a result of the bias correction. That is, thinking in terms of classic but non-robust mean squared error $\text{MSE}[\hat{\theta}] = \mathbb{E}[(\hat{\theta}-\theta)^2] = \text{var}(\hat{\theta}) +(\text{Bias}[\hat{\theta}])^2$, it is clear that a good estimator is a combination of both qualities. Classic goal has usually been to find out the minimum variance unbiased estimator (MVUE), which clearly minimizes MSE among the unbiased estimators. For more information about the iterative bootstrap, we refer the reader to the book by Hall \cite{hall2013bootstrap}.

We calculate the uncertainty estimates by ordering the sample obtained from the mother algorithm and calculate the vector component (pointwise) $1-2\alpha$ \textit{percentile intervals} at level $\alpha$ as $[\mathbf{\hat{f}}_{\star/0,\alpha}^*, \mathbf{\hat{f}}_{\star/0,1-\alpha}^*]$, where $\star/0$ means we consider both bias corrected ($\star$) and uncorrected (0) estimates and their intervals. As another option, the so-called \textit{basic bootstrap intervals} would be obtained with $[2\mathbf{\hat{f}} - \mathbf{\hat{f}}_{\star/0,{1-\alpha}}^*, 2\mathbf{\hat{f}} - \mathbf{\hat{f}}_{\star/0,1-\alpha}^*]$, where $\mathbf{\hat{f}}$ is the estimate obtained directly running \textsc{Kisu}. Basic bootstrap intervals have clearly a possibility of flipping the intervals upside down, which we observed also in the simulations. One could also go further and estimate the spectrum global uncertainty coverage, not just point by point local one. For more information, see \cite{efron1994introduction}.

\section{Simulations}
\label{sec:simulations}

\begin{figure}[h]
\begin{center}
\includegraphics[width=140mm]{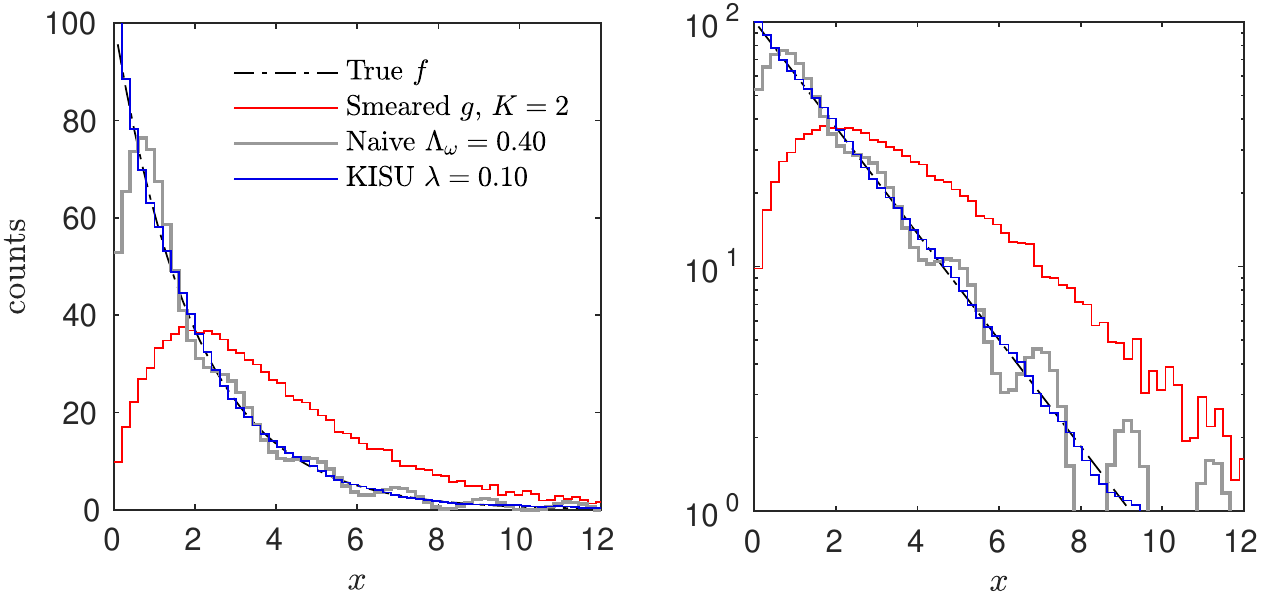}
\end{center}
\caption{Simplified fixed $K$ autoconvolution inverse using the naive Fourier domain inverse with spectral cut-off and \textsc{Kisu} based inverse. The true distribution is $f(x) = 1/\alpha \exp(-x/\alpha)$ with $\alpha = 2$.}
\label{fig:simplesimu1}
\end{figure}

\begin{figure}[h]
\begin{center}
$\begin{array}{ll}
\includegraphics[width=70mm]{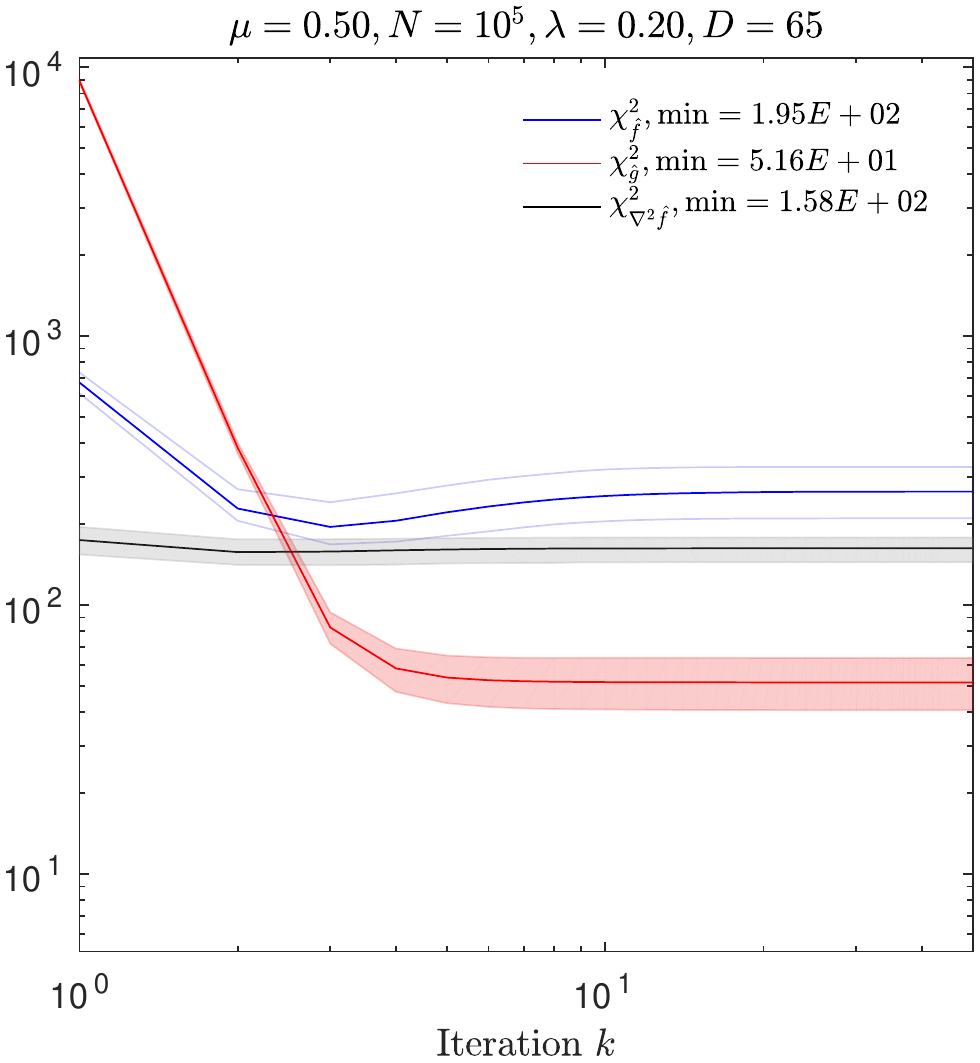} & \includegraphics[width=70mm]{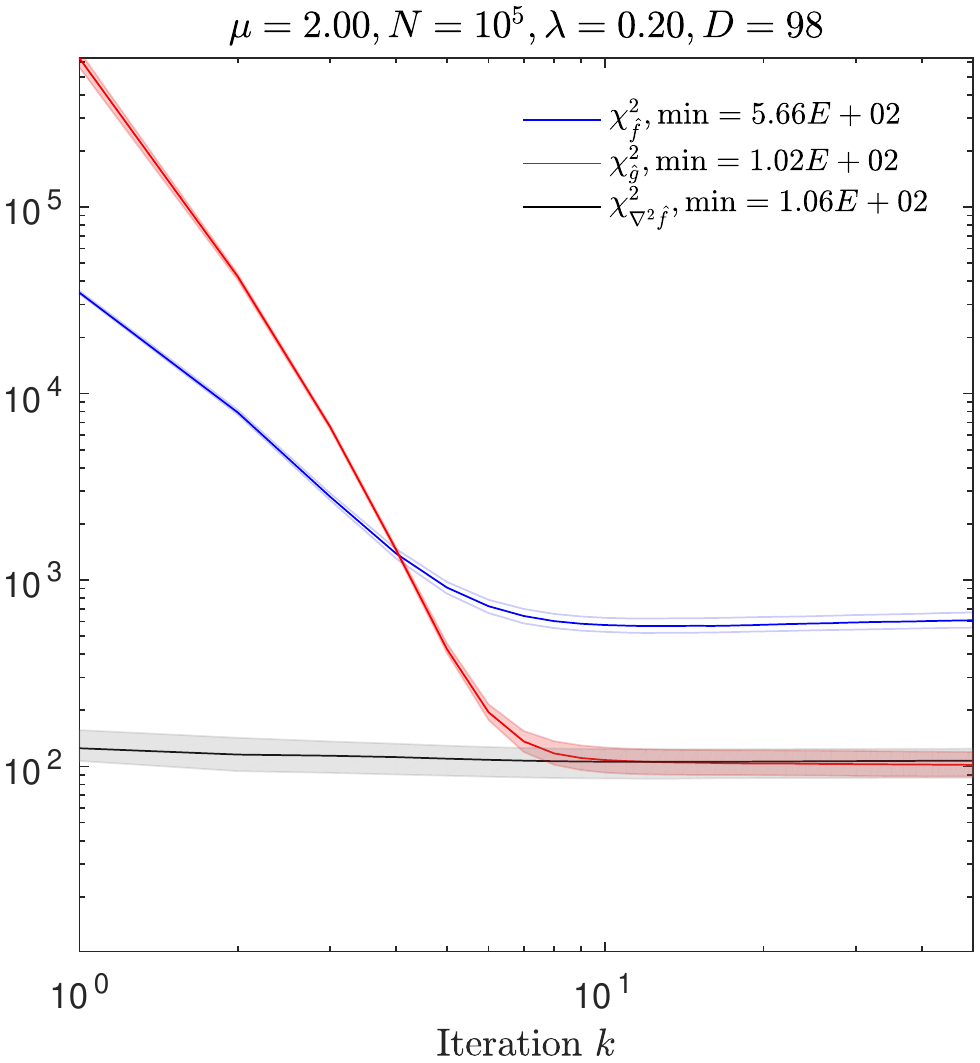}
\end{array}$
\end{center}
\vspace{-1em}
\caption{Typical \textsc{Kisu} algorithm iteration trajectories. Red lines denote the re-projection error, blue lines are the error with respect to the true distribution (available only in simulation) and black lines denote the variational regularity (smoothness).}
\label{fig:simiterationtrajectories}
\end{figure}

The first scenario to study is the the simplest fixed $K$ autoconvolution inverse of the negative exponential distribution without any uncertainty estimation. We illustrate this in Figure \ref{fig:simplesimu1} and compare with the naive Fourier domain inverse with spectral cut-off regularization, that is, we set high frequencies above the cut-off $\Lambda_\omega$ to zero, with $\omega \in [0,1]$. The spectral cut-off is tuned to give the best performance for this scenario. We observe nearly perfect reconstruction with \textsc{Kisu} and very large oscillations with the naive Fourier domain algorithm. These oscillations seemed to appear immediately after a small amount of counting fluctuations in the observed spectrum $g$, which we simply added here as a Gaussian noise. For a typical iteration trajectories, see Figure \ref{fig:simiterationtrajectories}. We see that the variational regularity cost is kept as almost constant, and the error with respect to the ground truth and the re-projection error behave in a same way, which is always called for behavior.

\begin{figure}[h]
\begin{center}
\includegraphics[width=80mm]{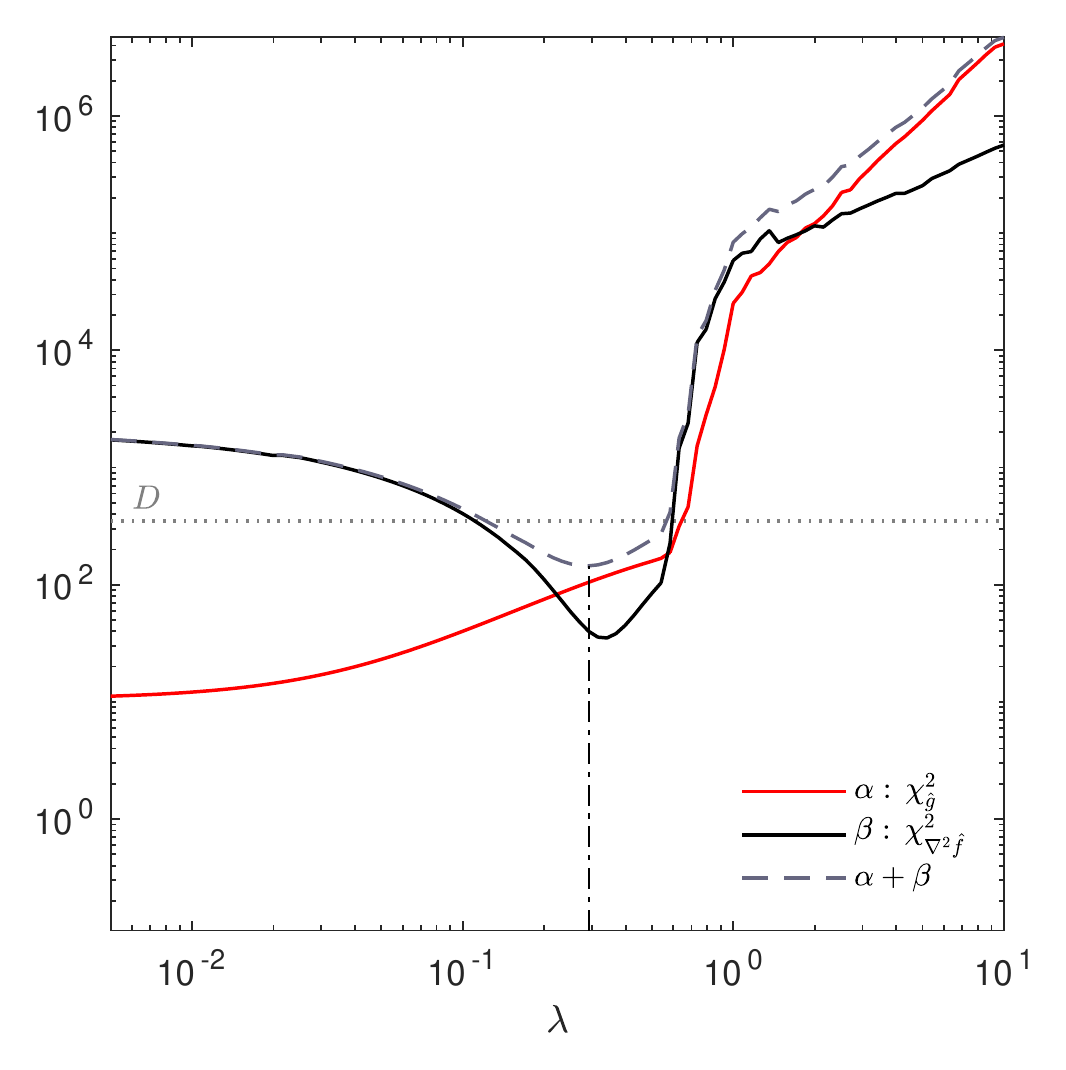}
\end{center}
\vspace{-1em}
\caption{Data driven equilibrium between the re-projection cost $\alpha$ and the variational regularity cost $\beta$. The horizontal line denotes the number of bins $D$ in the measured histogram $\sim$ number of degrees of freedom. }
\label{fig:regularization}
\end{figure}

We tried to two different `data-driven' ways of choosing the regularization parameter, see Figure \ref{fig:regularization}. First, the regularization parameter $\lambda$ was selected as the equilibrium point $\min(\alpha + \beta)$, where the re-projection error is $\alpha = \chi^2_{\hat{g}} = ||(\mathbf{g} - \hat{\mathbf{g}})/\sqrt{\mathbf{g}}||_{\ell_2}^2$ and the regularity cost is $\beta = \chi^2_{\nabla^2\hat{f}} = ||\nabla^2 \hat{\mathbf{f}} / \sqrt{\hat{\mathbf{f}}} ||_{\ell_2}^2$, with operations taken element wise. Re-projection is simply defined as $\hat{g} = \mathcal{\hat{F}}(\hat{f})$, available after the inversion. The values for different $\lambda$ values are obtained via brute force loop. This was repeated for each simulation scenario separately. The second criteria was to use the Morozov like discrepancy principle point where the re-projection error $\alpha$ is approximately the same as the number of histogram bins $D$. For the rest of the simulations, we use the equilibrium solution. When $\lambda \rightarrow 1$, the solution becomes over smooth which results in a loss of high frequency details and when $\lambda \rightarrow 0$, we start to fit the noise in high frequencies. In general, slightly larger $\lambda$ values were always obtained as the equilibrium solution compared to the discrepancy principle.

\begin{figure}[H]
\begin{center}
$\begin{array}{ll}
\includegraphics[width=65mm]{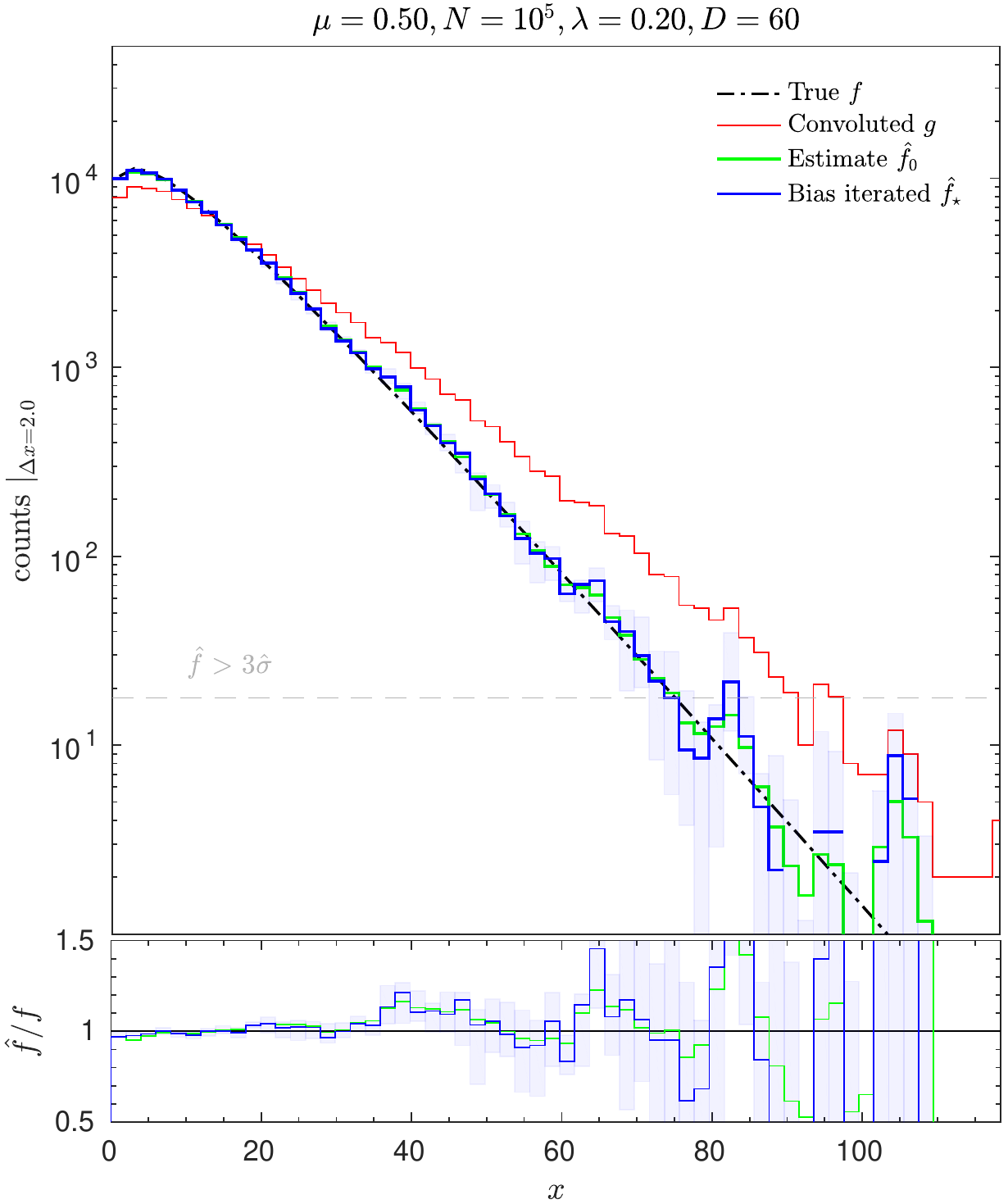} & \includegraphics[width=65mm]{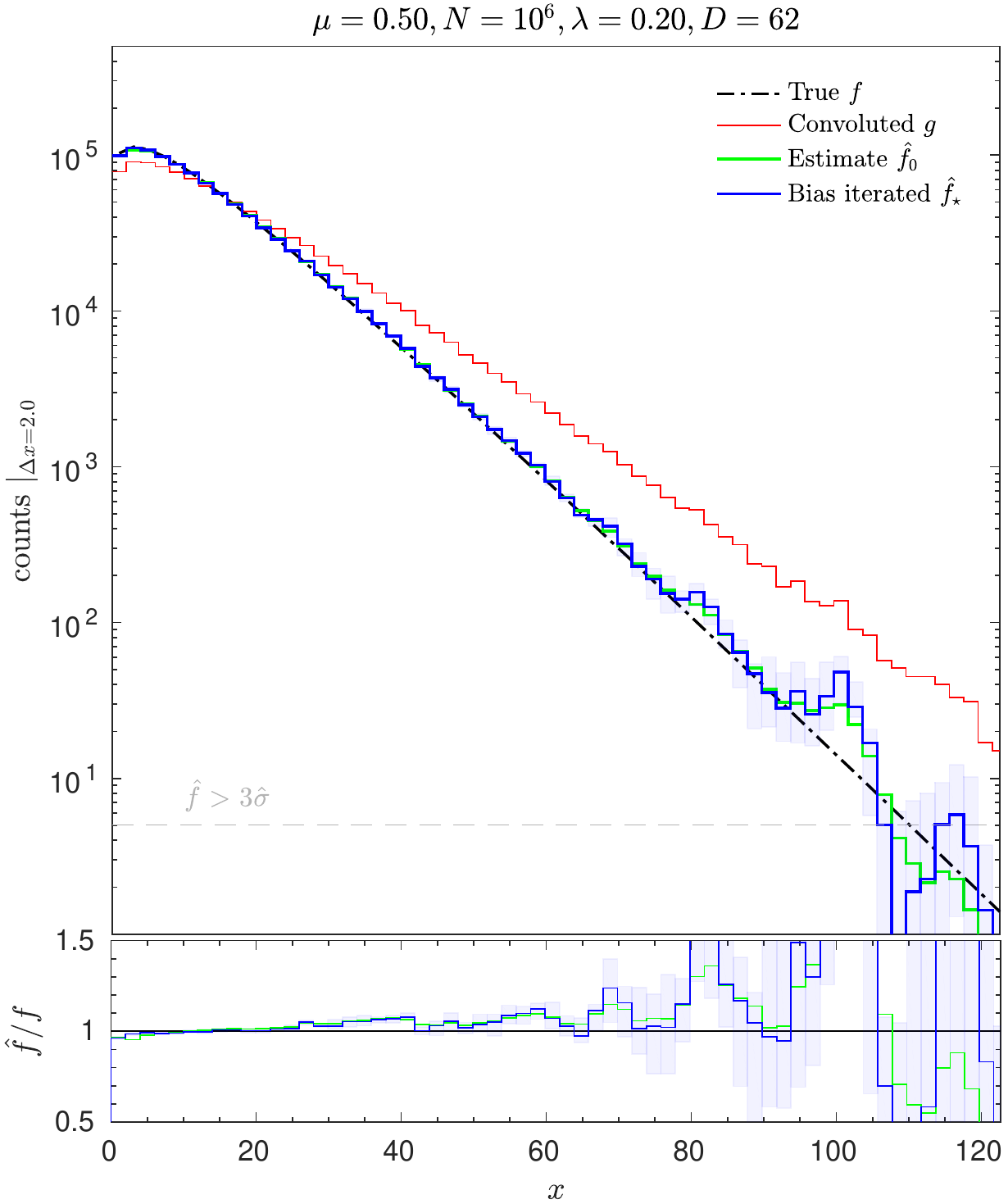} \\
\includegraphics[width=65mm]{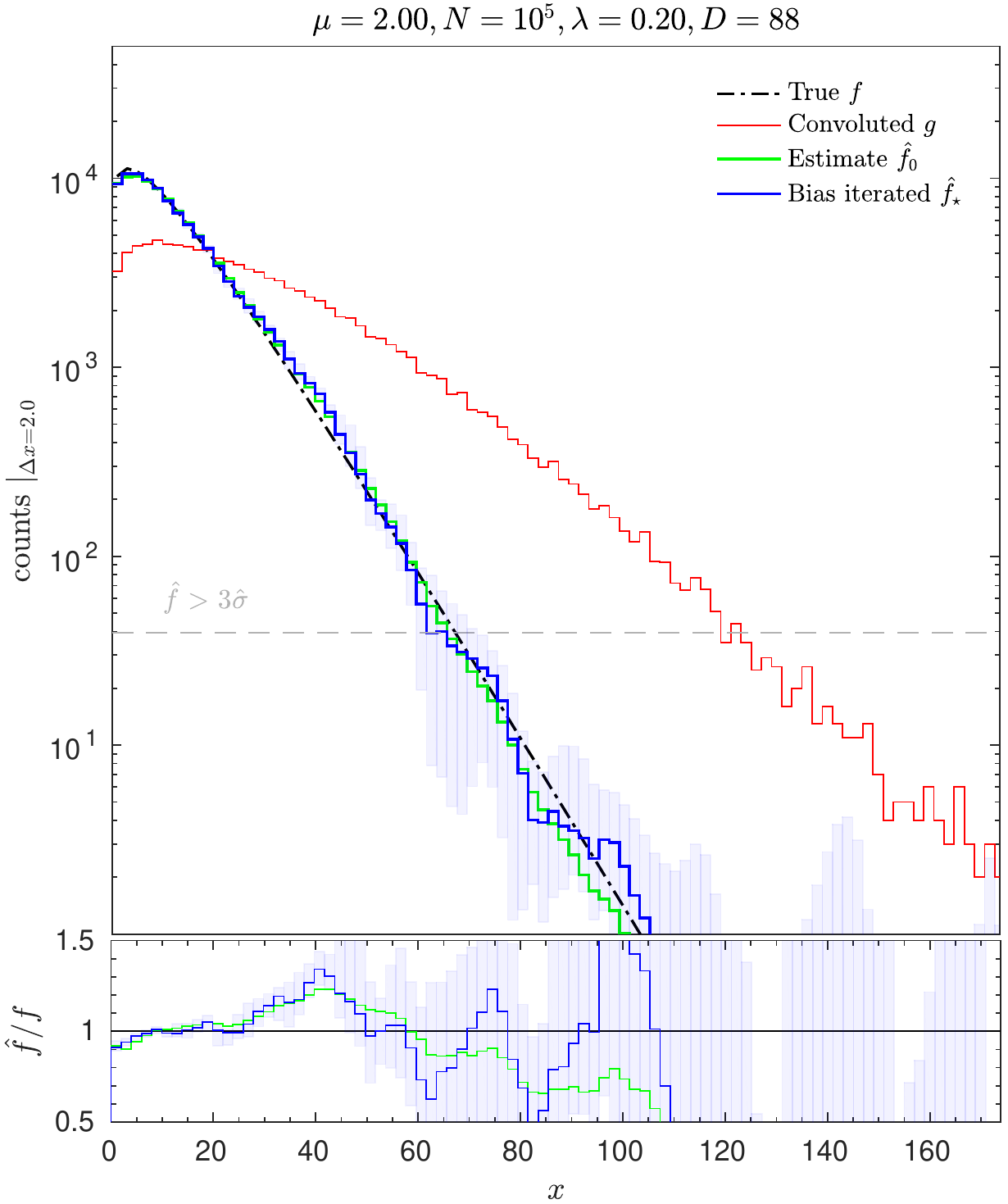} & \includegraphics[width=65mm]{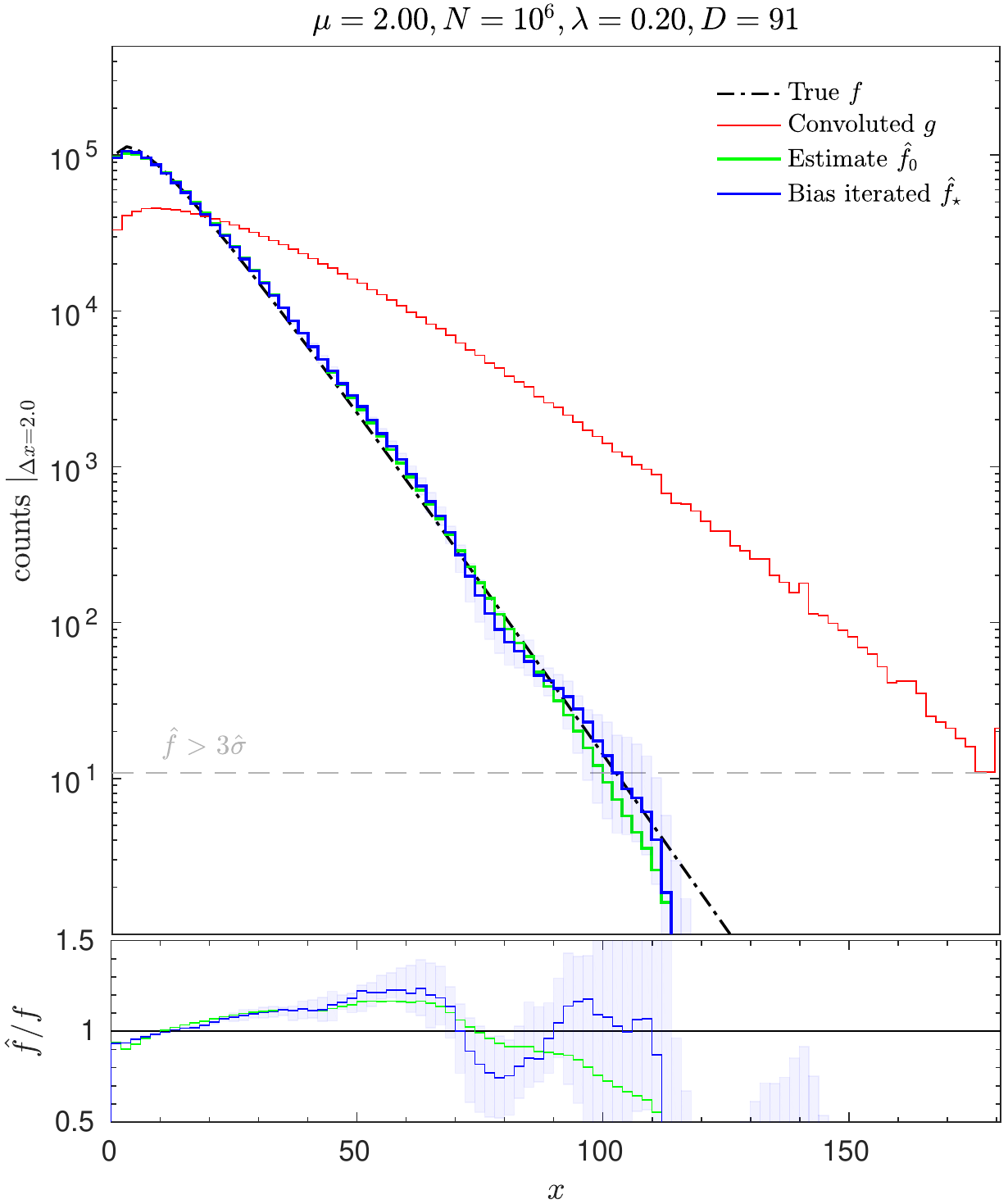} \\
\includegraphics[width=65mm]{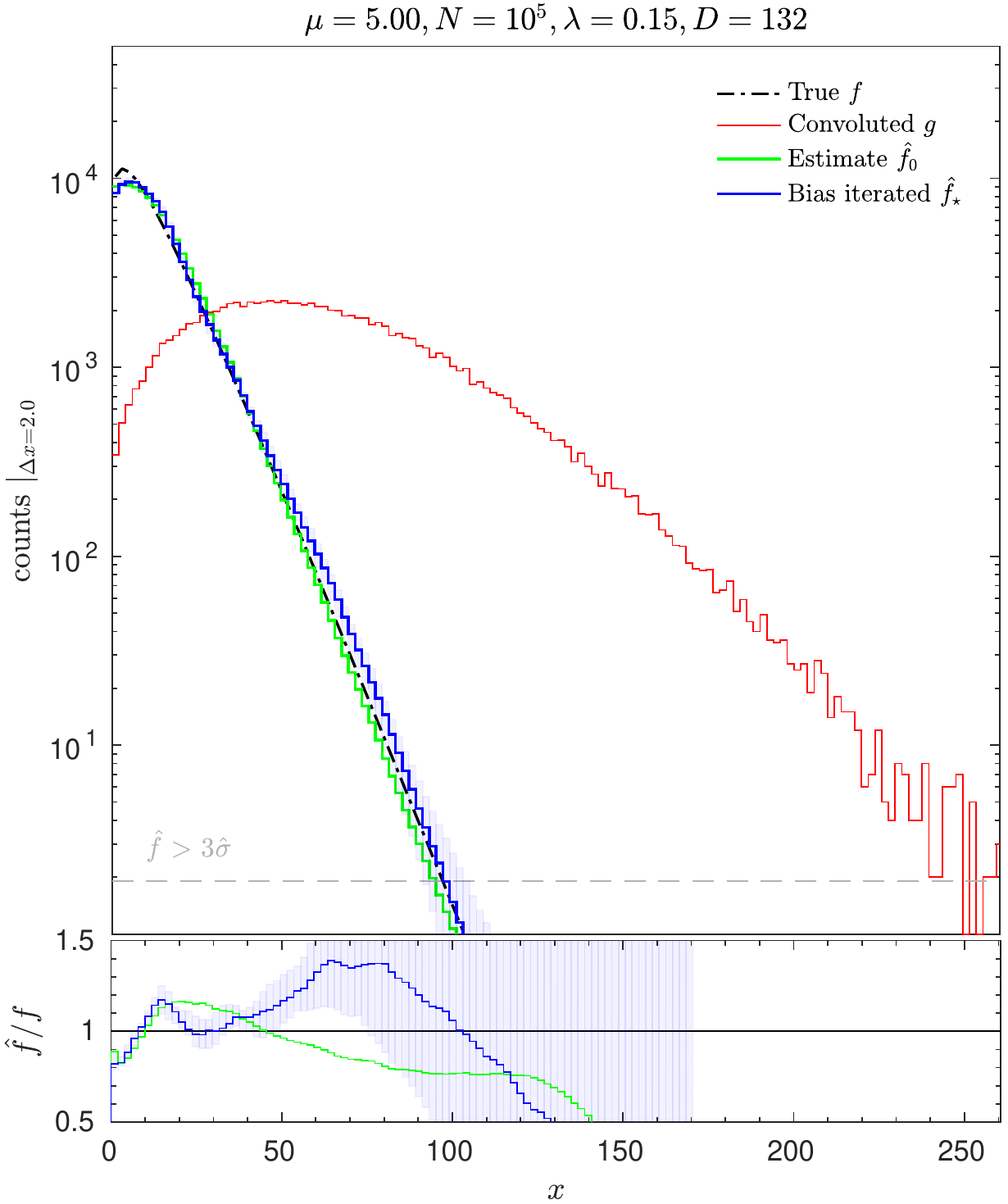} & \includegraphics[width=65mm]{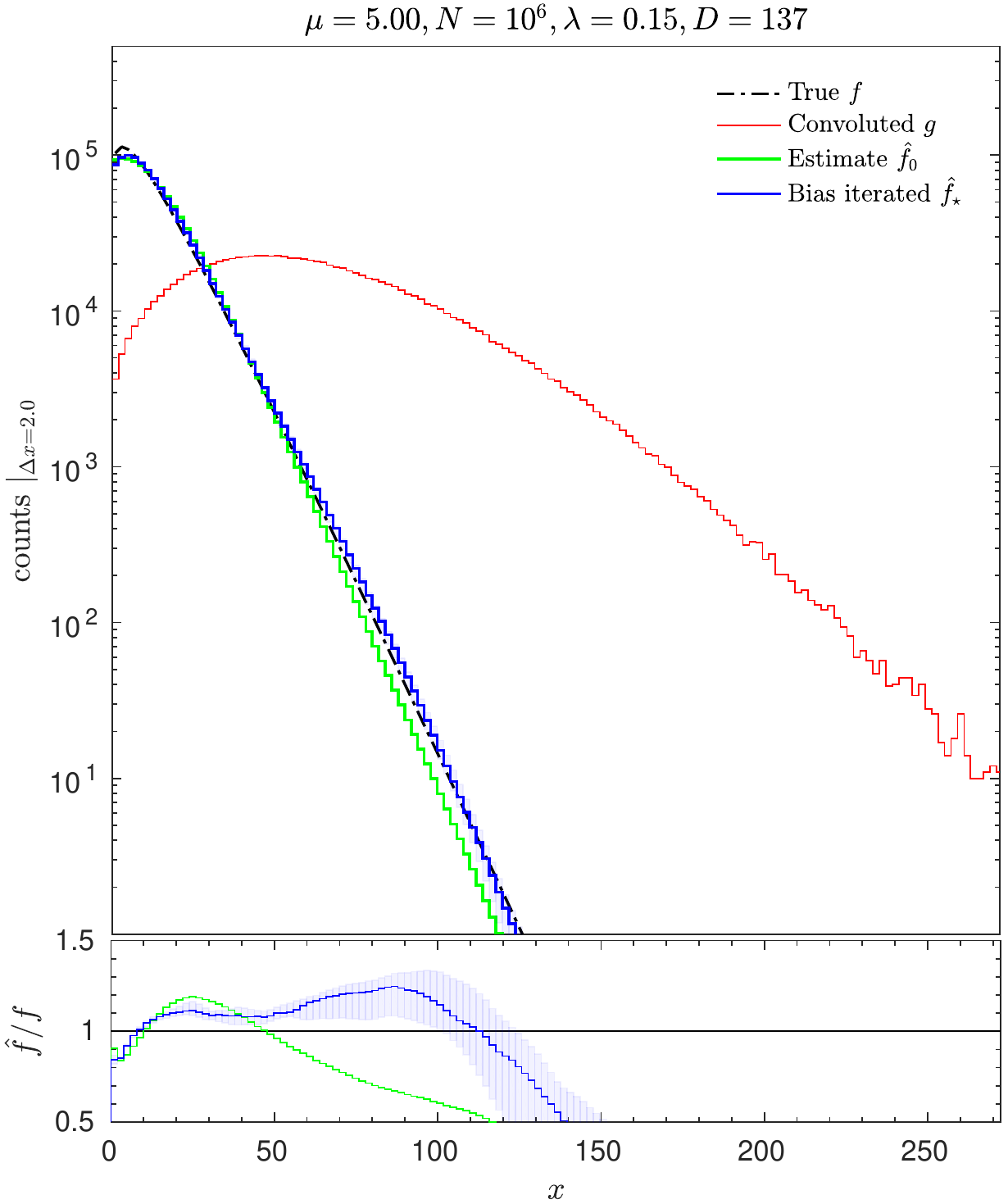} \\
\end{array}$
\end{center}
\caption{Simulations with $f(x) = P_{\text{NBD}}(x; \langle n \rangle = 12.5, k = 1.4)$. Bootstrap based uncertainties are denoted with blue. }
\label{fig:simulationNBD}
\end{figure}

\begin{figure}[H]
\begin{center}
$\begin{array}{ll}
\includegraphics[width=65mm]{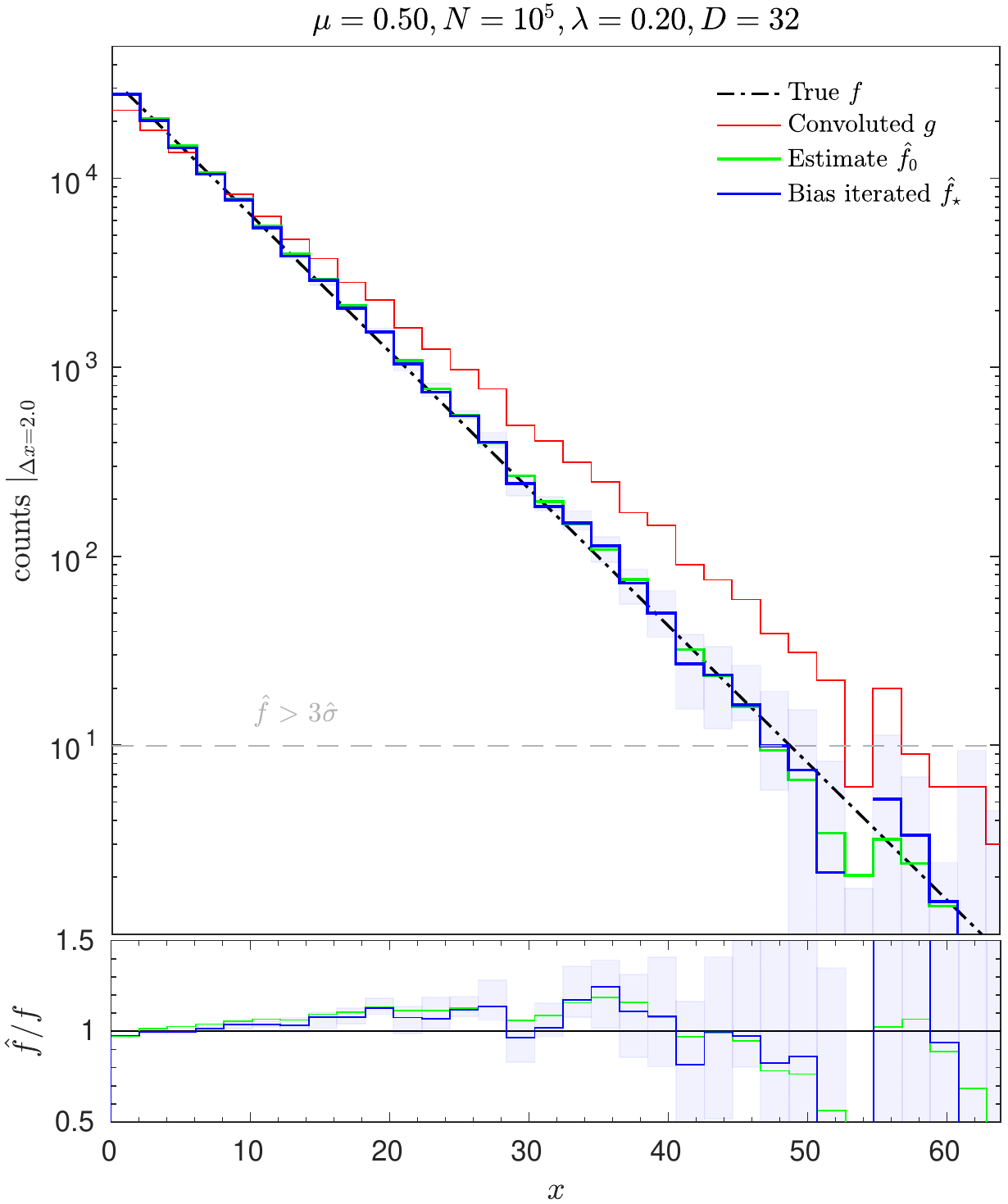} & \includegraphics[width=65mm]{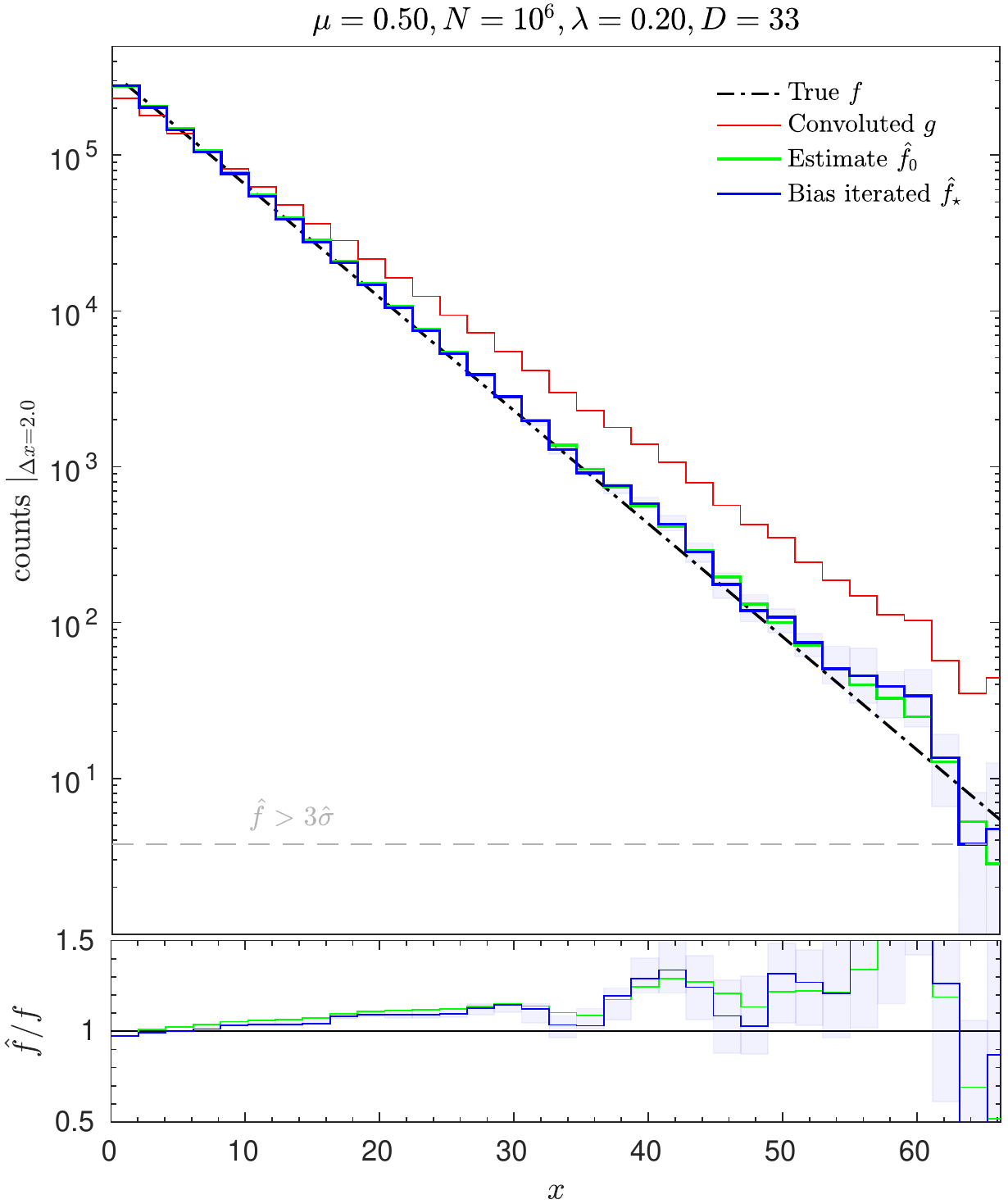} \\
\includegraphics[width=65mm]{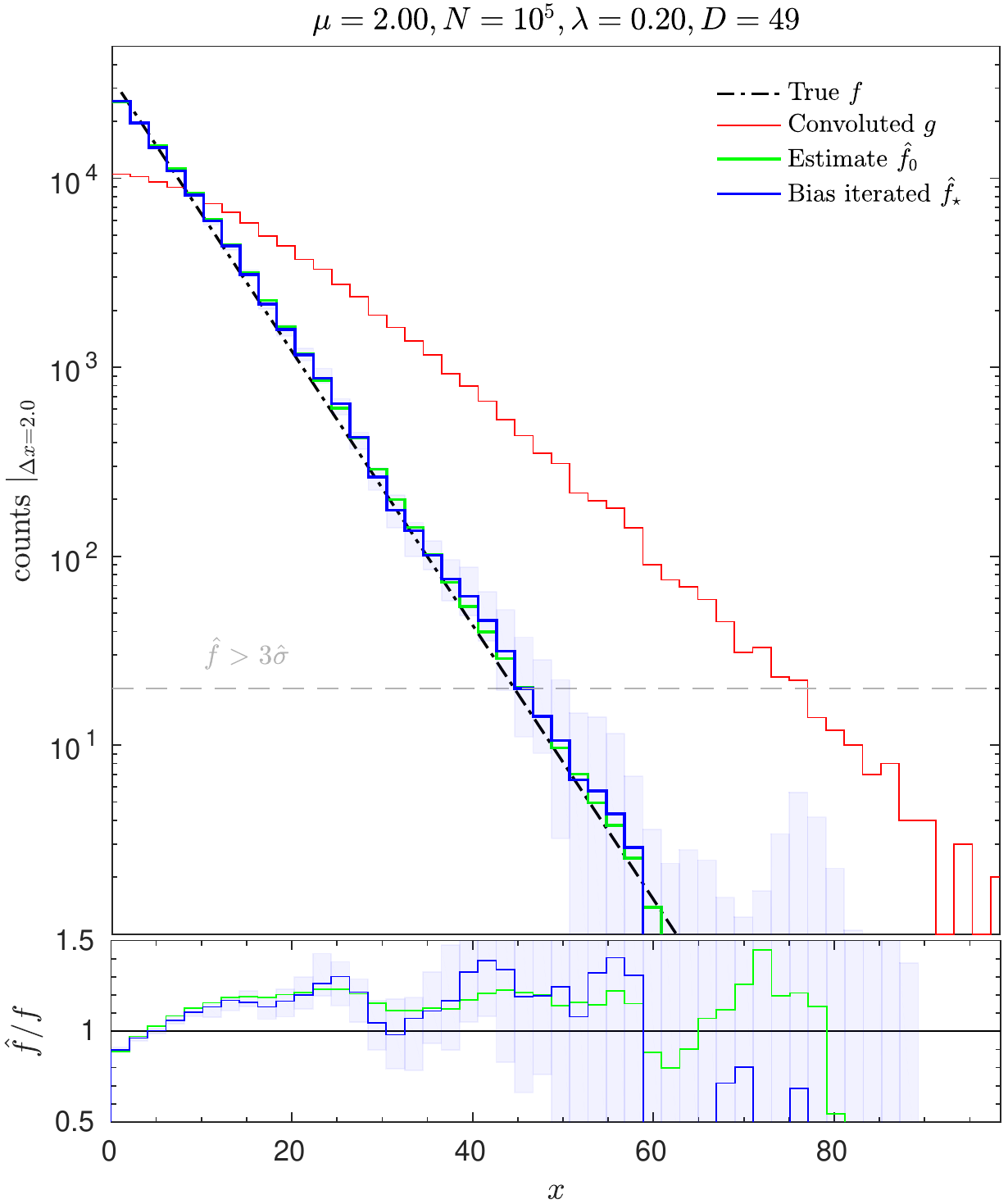} & \includegraphics[width=65mm]{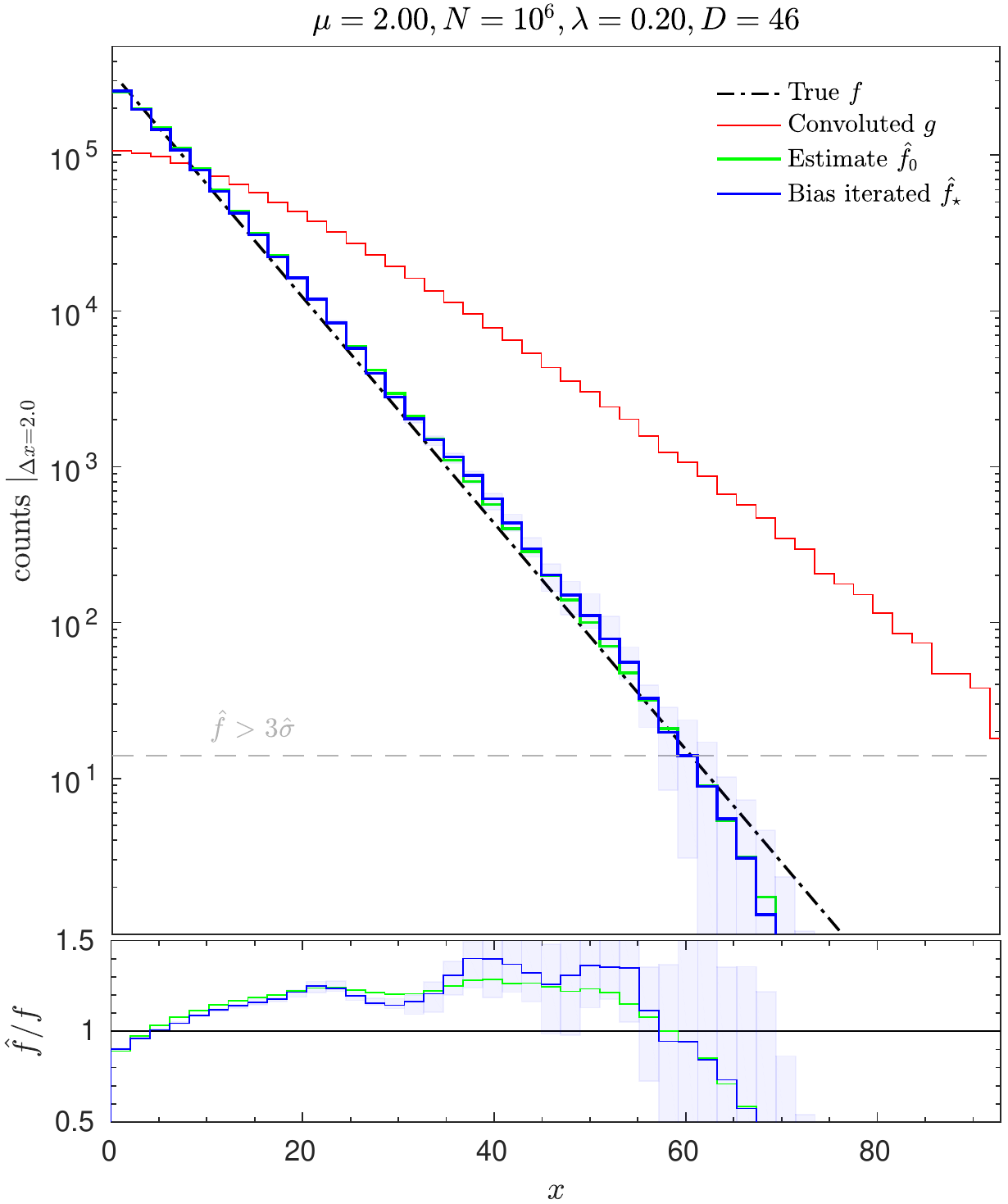} \\
\includegraphics[width=65mm]{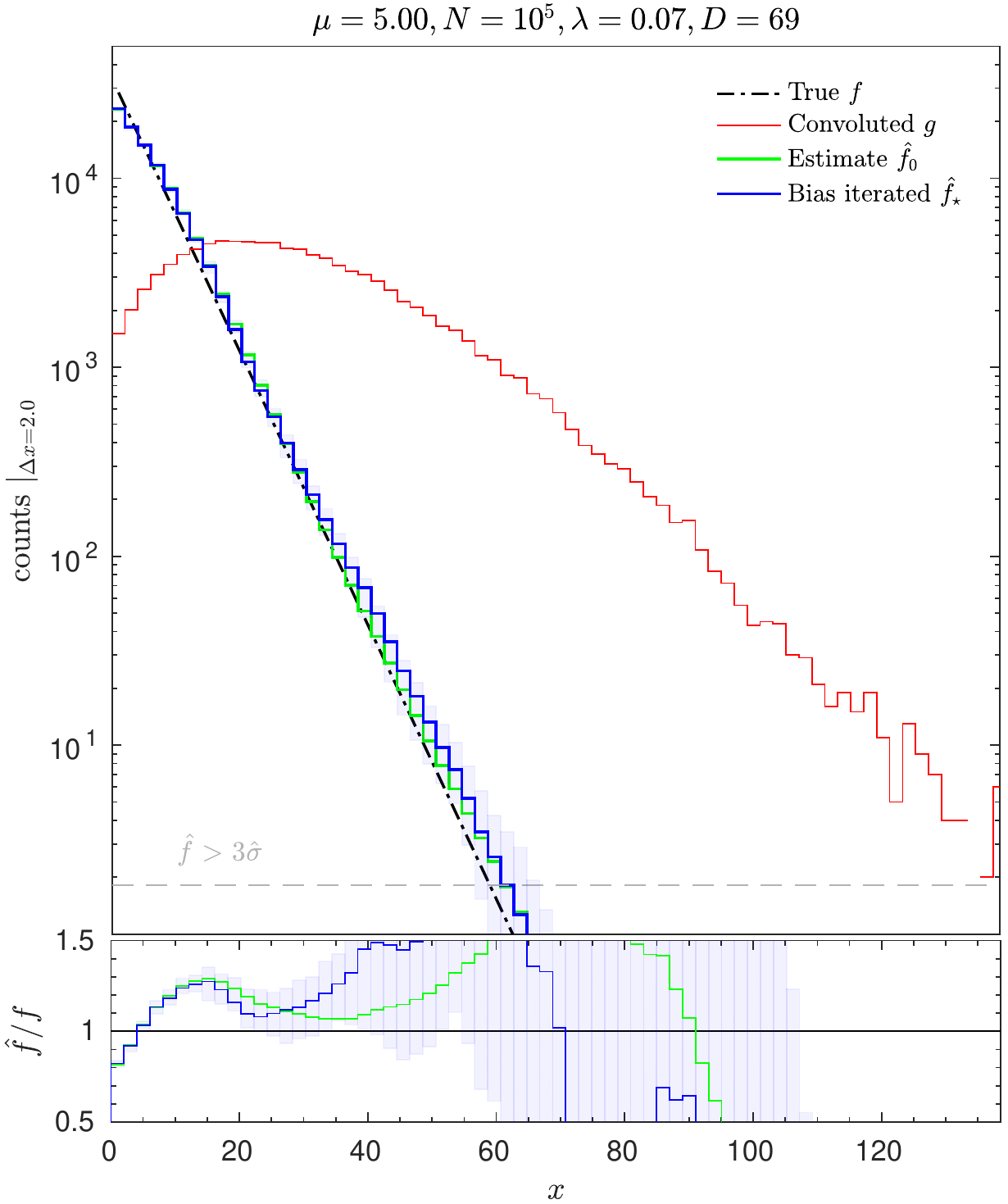} & \includegraphics[width=65mm]{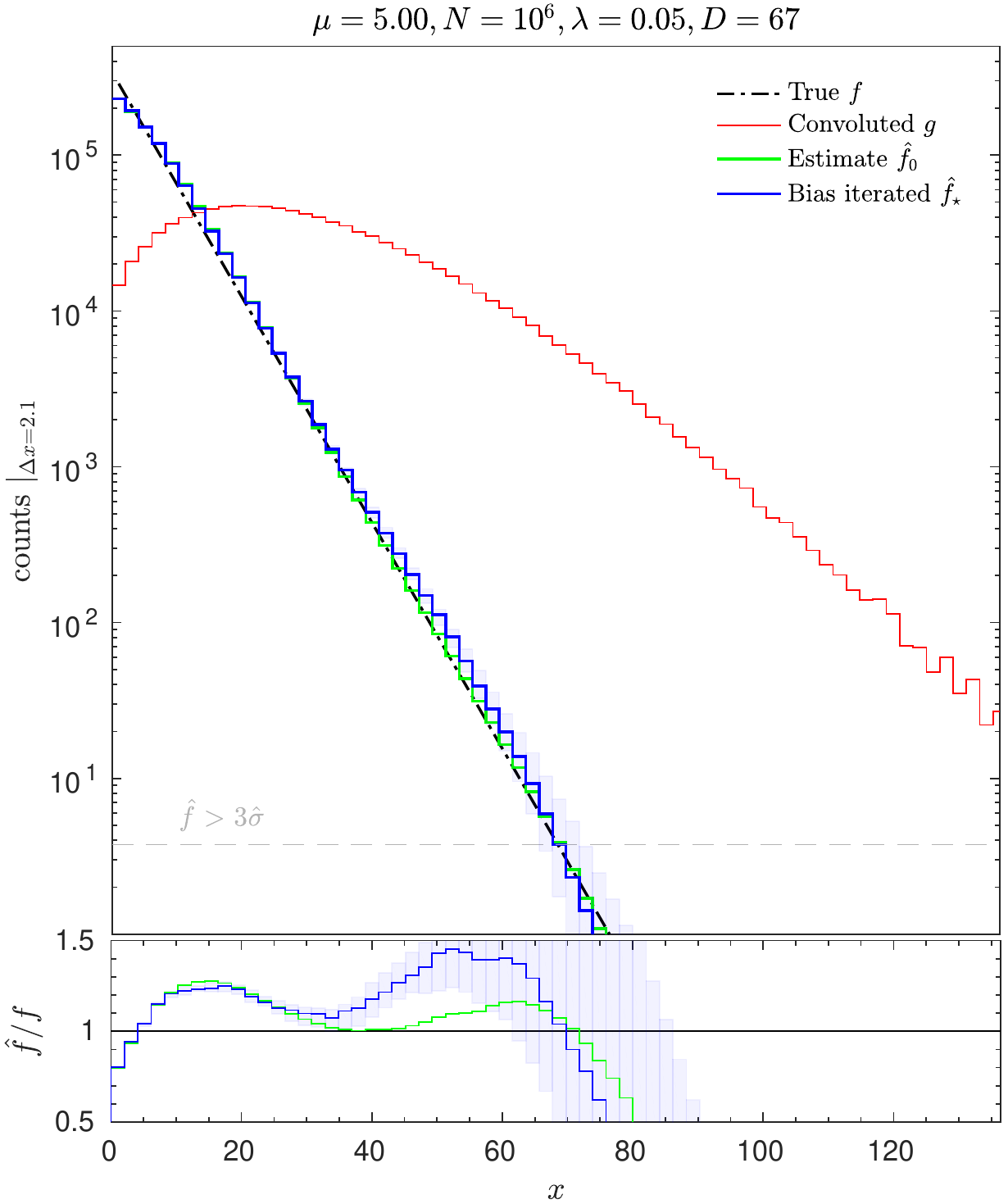} \\
\end{array}$
\end{center}
\caption{Simulations with $f(x) = 1/\alpha \exp(-x/\alpha)$ with $\alpha = 6$. Bootstrap based uncertainties are denoted with blue.}
\label{fig:simulationEXP}
\end{figure}

In more demanding simulations we have a scenario which corresponds approximately to inverting minimum bias pileup at the LHC. We chose fully analytical distributions as the ground truth $f$ and draw samples via von Neumann acceptance-rejection Monte Carlo sampling. The distribution is a negative binomial with parameters given in Figure \ref{fig:simulationNBD} and exponential in Figure \ref{fig:simulationEXP}. The number of events $N$ and the Poisson $\mu$ were varied, with the histogram bin size $\Delta x$ kept approximately fixed. One must remember that the overall statistics scales also with the $\mu$-value, so there must be an optimal point with respect to the pileup deterioration and the collected number of events.

Convergence of the iterative algorithms was obtained in every scenario, and the fundamental limitations were hit when the smeared distribution $g$ was almost purely Gaussian, which was the case when $\mu \sim 10$ or more. In that case, the inverted distribution was lacking any fine structure. Frequentist point wise uncertainty bands $(\alpha = 0.05)$ $\sim 95$ CL were obtained from the bootstrap procedures. The coverage is heuristically reasonably when comparing with respect to the truth, however, the nonlinear oscillation is not clearly covered by the bootstrap. A $3\sigma$ signal to uncertainty ratio, shown in each figure, seems to be a reasonable sanity threshold cut.

In Figure \ref{fig:simulationNBD}, we observe some oscillation or `nucleation' of certain eigenmodes in the solution, which is not an unusual phenomena in inverse problems. We see that the bias correction seems to increase variance of the estimates slightly, which is visible in bin-by-bin fluctuations. This is expected due the to bias-variance tradeoff. The optimal number of bias iterations should be studied further, because it is a free parameter, unless taken till convergence. In the NBD simulation case, bias correction seems to have amplified slightly the oscillation via recursion. Here, we used three iterations. Thus, it can behave as an additional regulator of the problem. The size of bootstrap samples for the daughter and mother algorithms should be taken as large as computational resources allow. As a practical guideline, extensive simulations should be always used to investigate the problem specific stability, regularization, bias-variance and uniqueness.

\section{LHC data inversion}
\label{sec:LHCdata}

For this proof-of-concept study, we use ALICE proton-proton charged particle multiplicity spectra measured at $\sqrt{s} = 0.9$ and 7 TeV from  \href{https://www.hepdata.net/record/ins1614477}{HEPData} \cite{ALICE:2017pcy}. The publication includes also data at $\sqrt{s}=8$ TeV, which we leave out here because of very similarity with $\sqrt{s} = 7$ TeV, but we include it in our analysis code available online. Phenomenologically, the average central multiplicity density $dN_{ch}/d\eta$ in proton-proton follows Regge like power law scaling $\propto s^{\alpha(0)-1}$ with the effective Pomeron intercept $\alpha(0) ~ \simeq 1.1$. The event selection definitions are: the INEL (minimum bias inelastic) which in this case means minimal activity in any of the trigger subdetectors and the NSD (non-single-diffractive), which is event rapidity topology based selection. The idea behind the NSD is to suppress \textit{qualitatively} single diffraction events within forward system mass range which result in a large enough pseudorapidity gap on forward or backward pseudorapidity side of the detector. The single diffractive suppression is done simply by requiring event activity on both forward and backward triggers. In addition, the INEL events may have either $N_{ch} \geq 0$ or $\geq 1$ particles (tracklets) required in the central region $|\eta|<1$. We use $N_{ch} \geq 0$ class, because diffractive events can be with $N_{ch} = 0$ at central but trigger forward detectors.

\begin{figure}[H]
\begin{center}
$\begin{array}{ll}
\includegraphics[width=68mm]{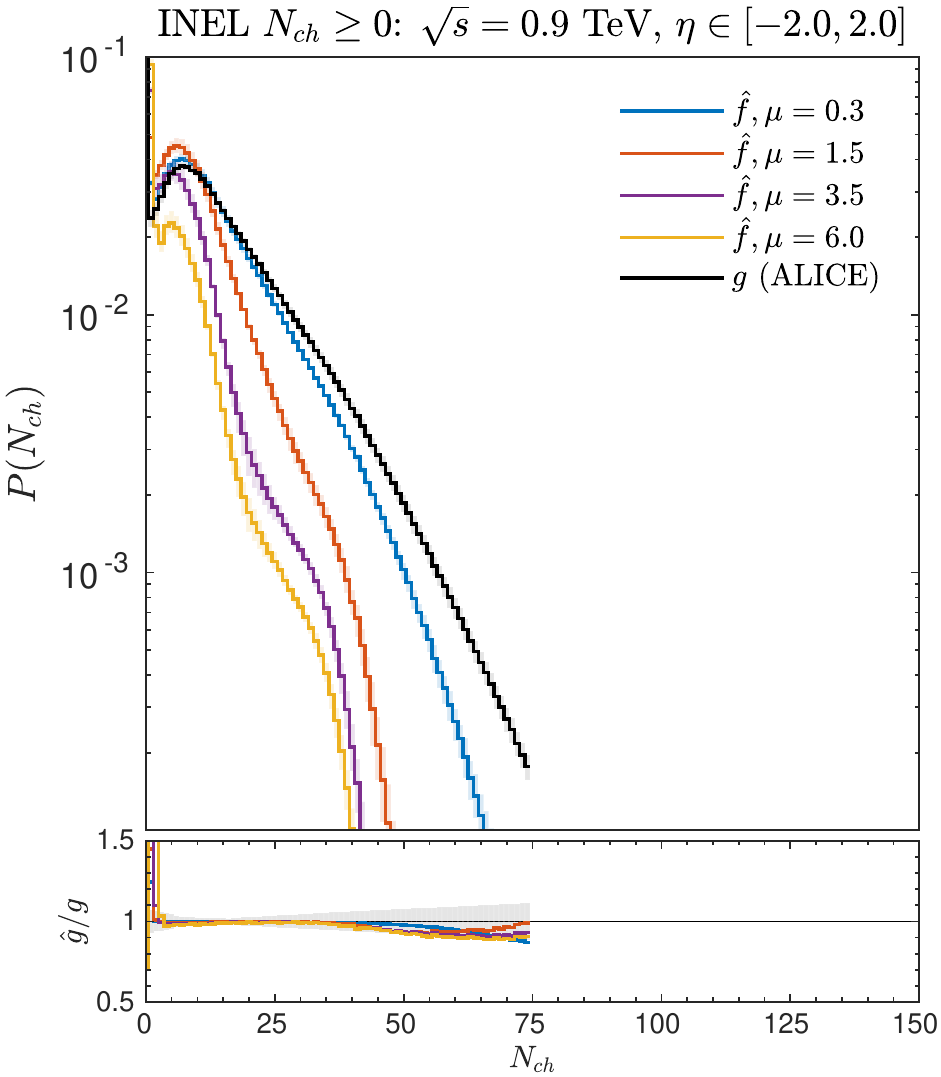} &
\includegraphics[width=68mm]{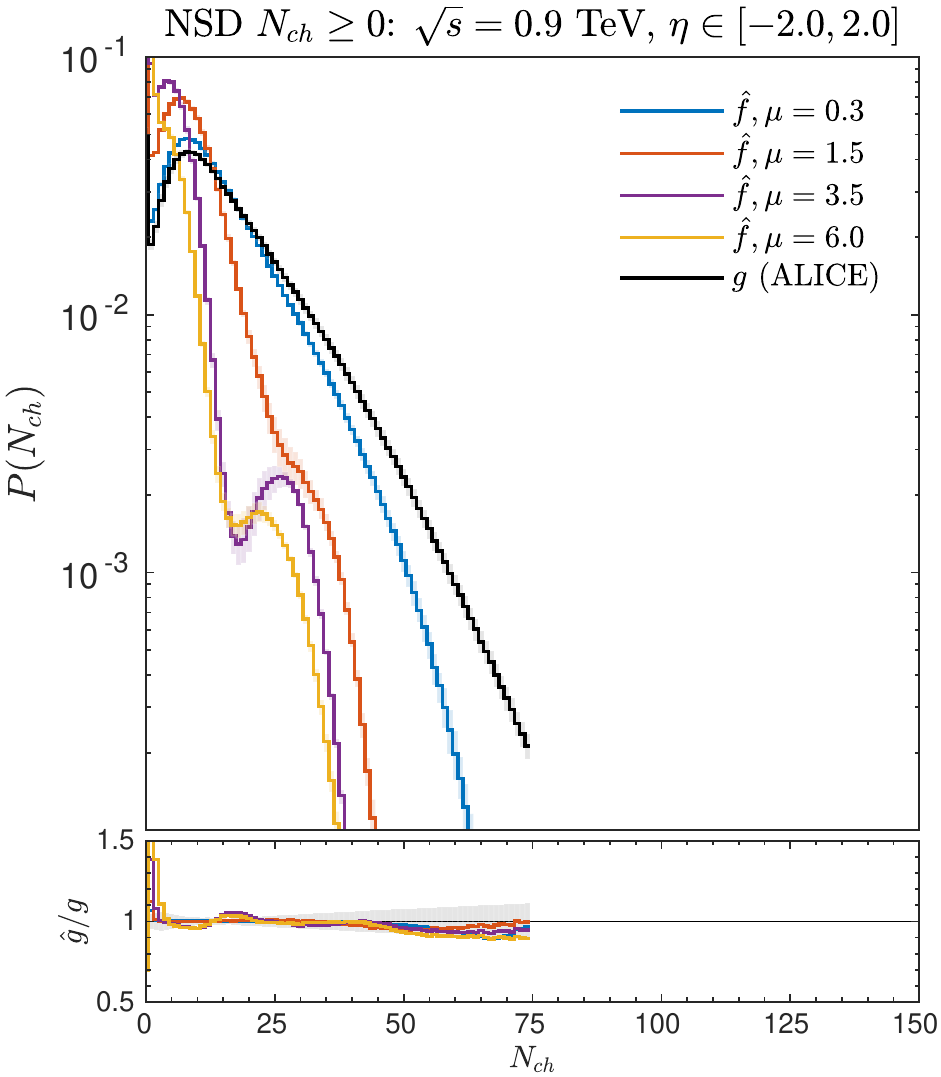} \\
\includegraphics[width=68mm]{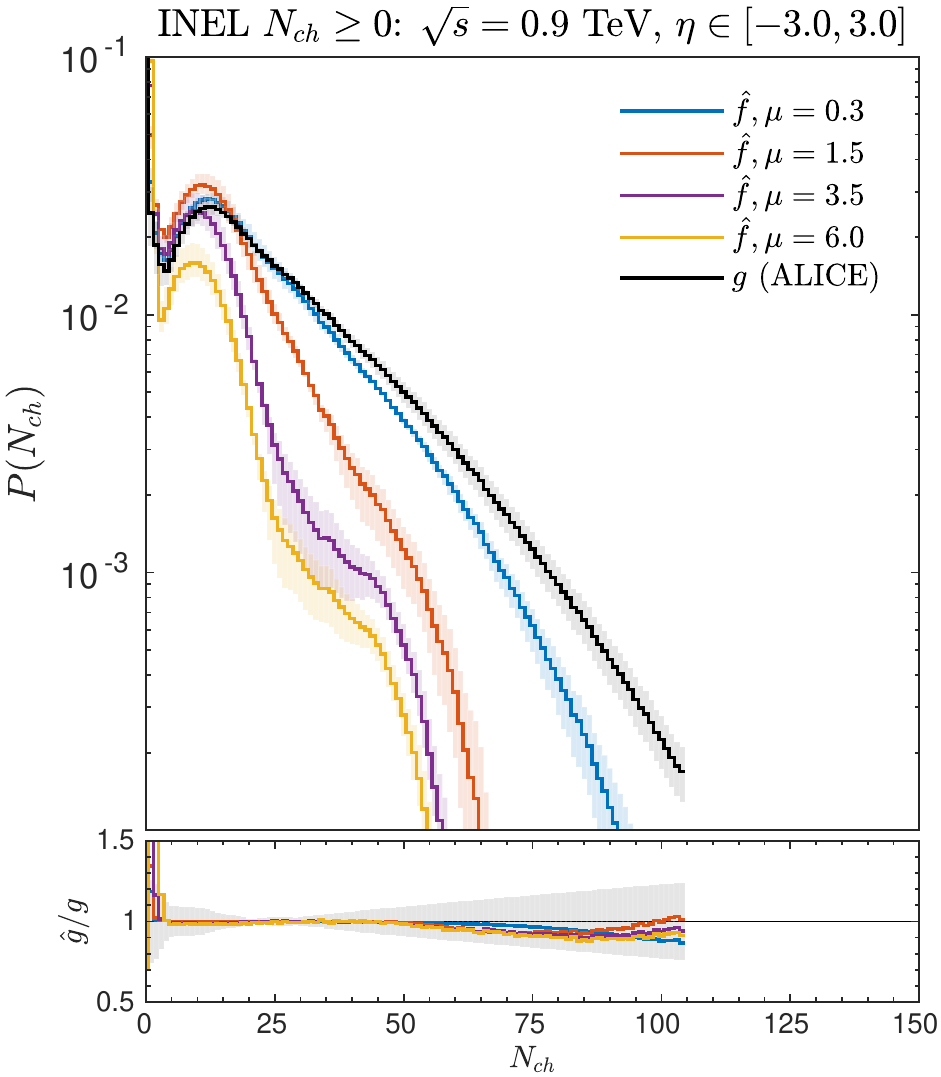} &
\includegraphics[width=68mm]{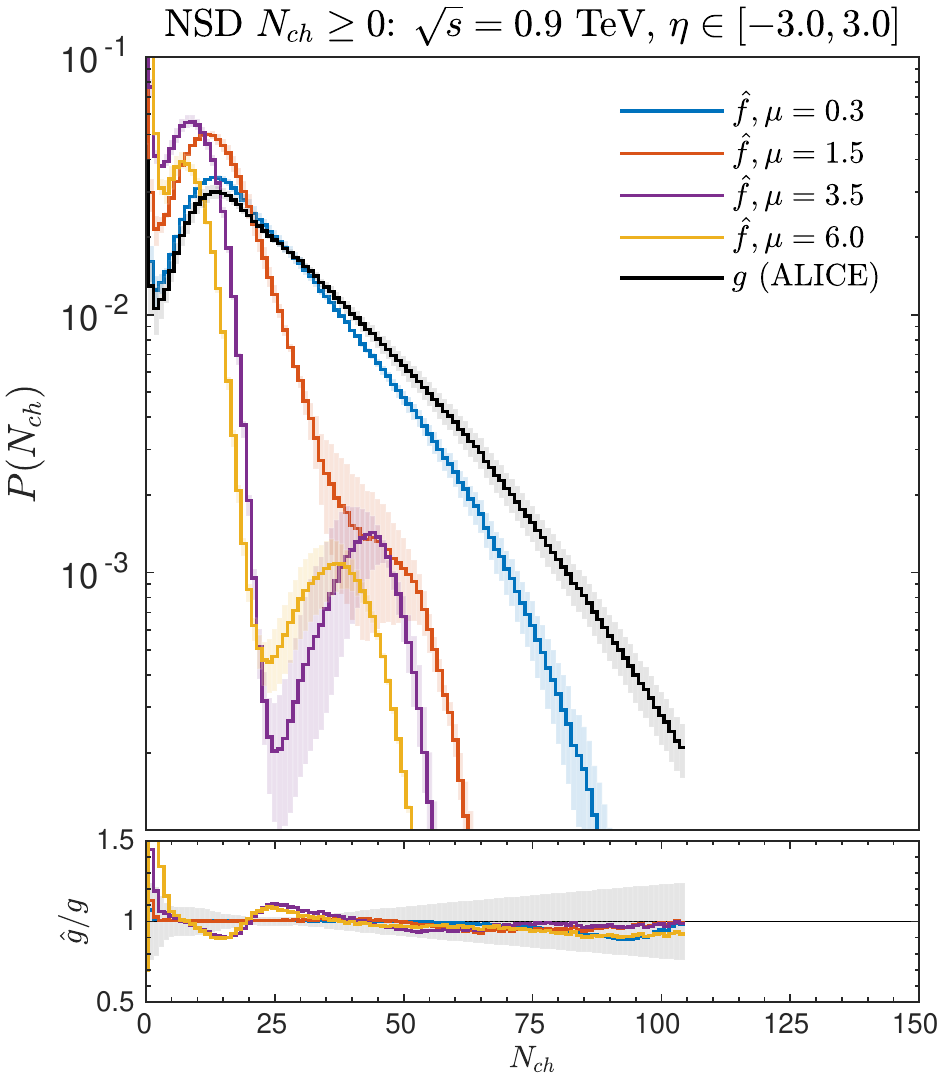} \\
\includegraphics[width=68mm]{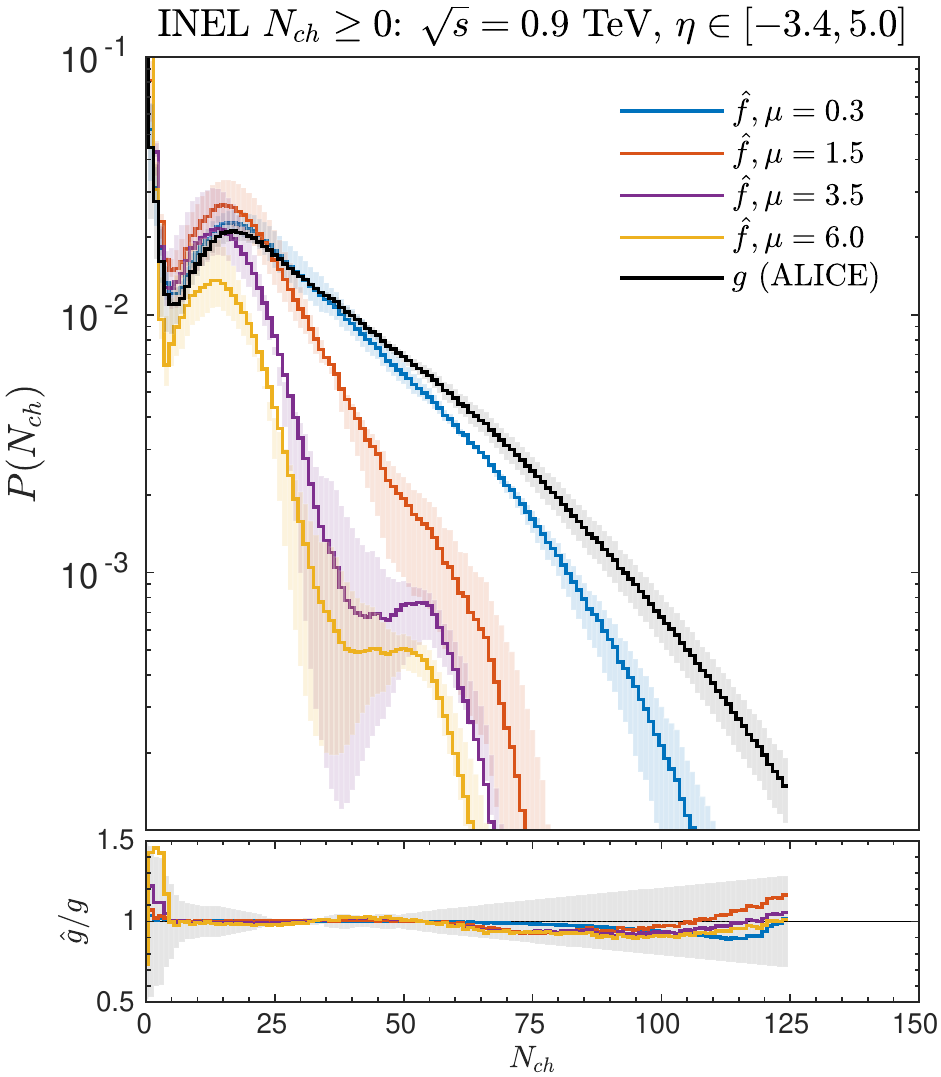} & 
\includegraphics[width=68mm]{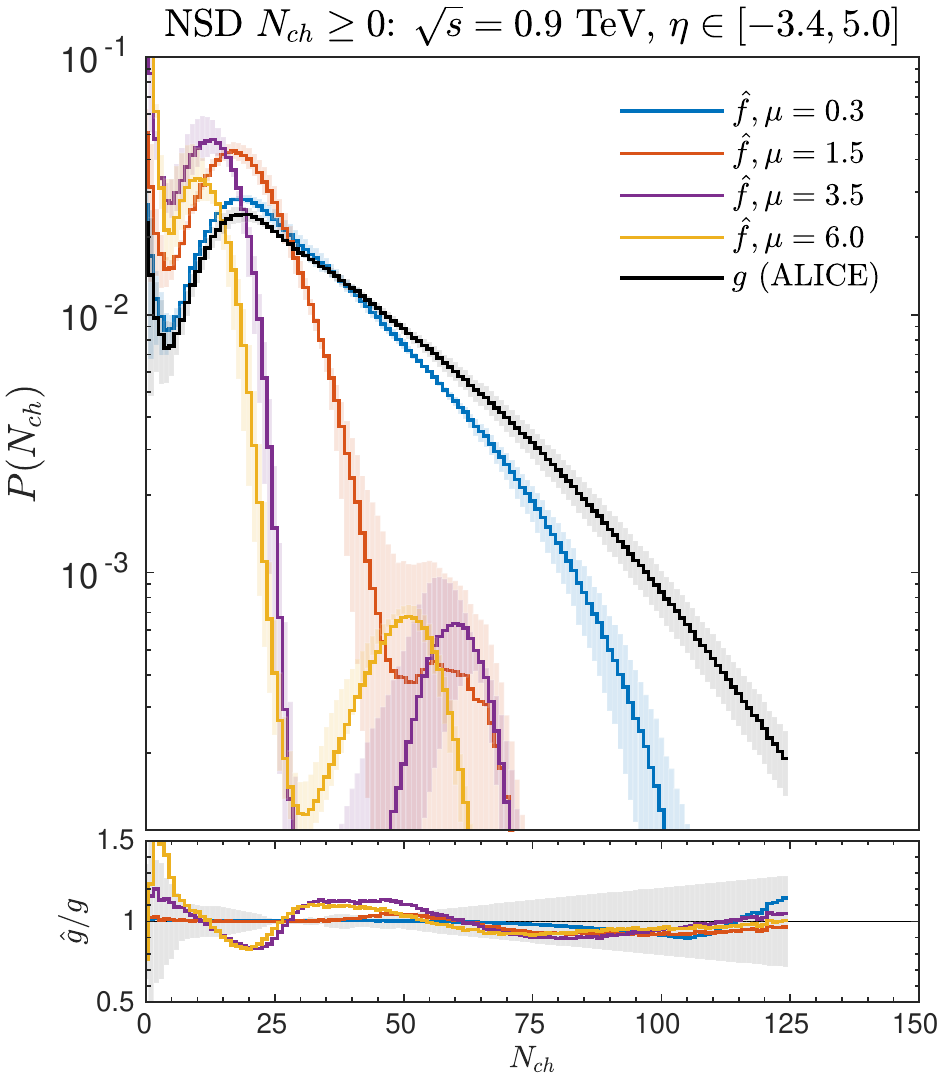} \\
\end{array}$
\end{center}
\caption{ALICE INEL (left) and NSD (right) data at $\sqrt{s} = 0.9$ TeV \cite{ALICE:2017pcy} for three different pseudorapidity intervals and the \textsc{Kisu} inversion results under different $\mu$-hypothesis.}
\label{fig:ALICE1}
\end{figure}

\begin{figure}[H]
\begin{center}
$\begin{array}{ll}
\includegraphics[width=68mm]{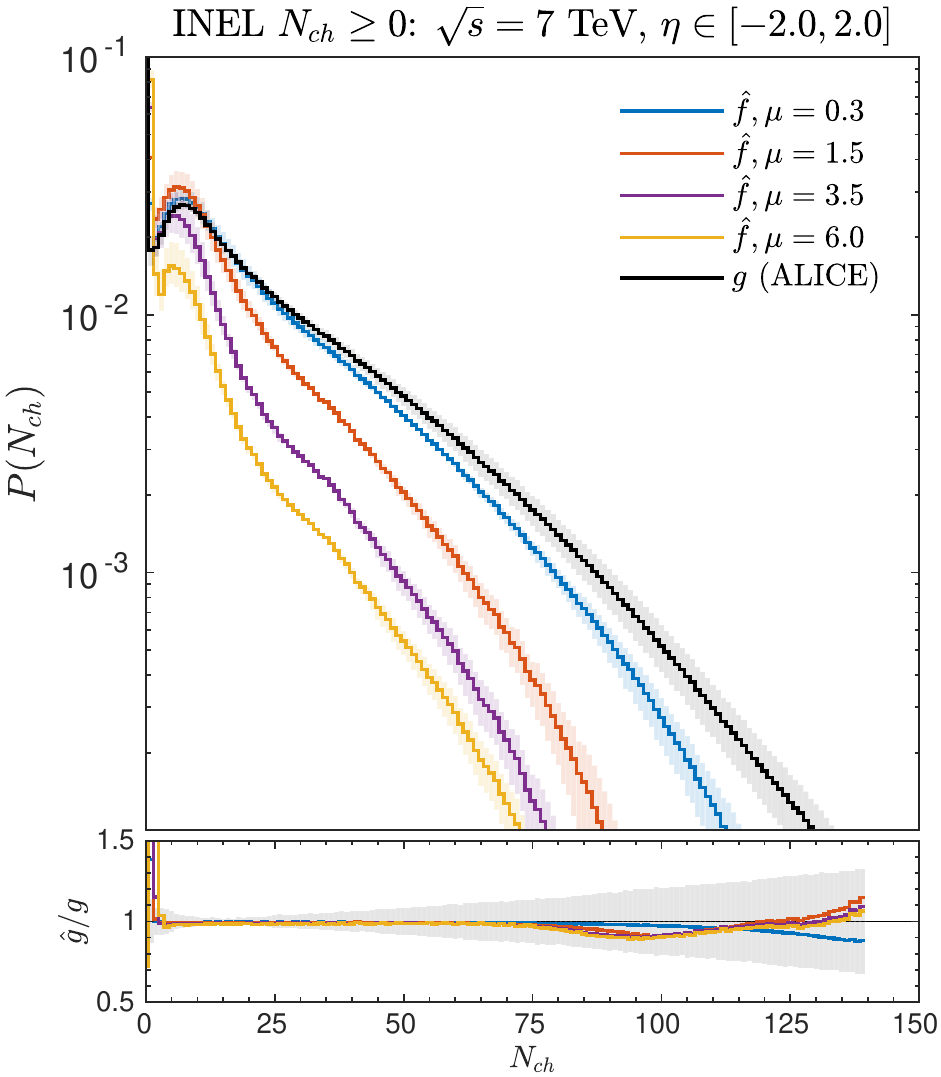} &
\includegraphics[width=68mm]{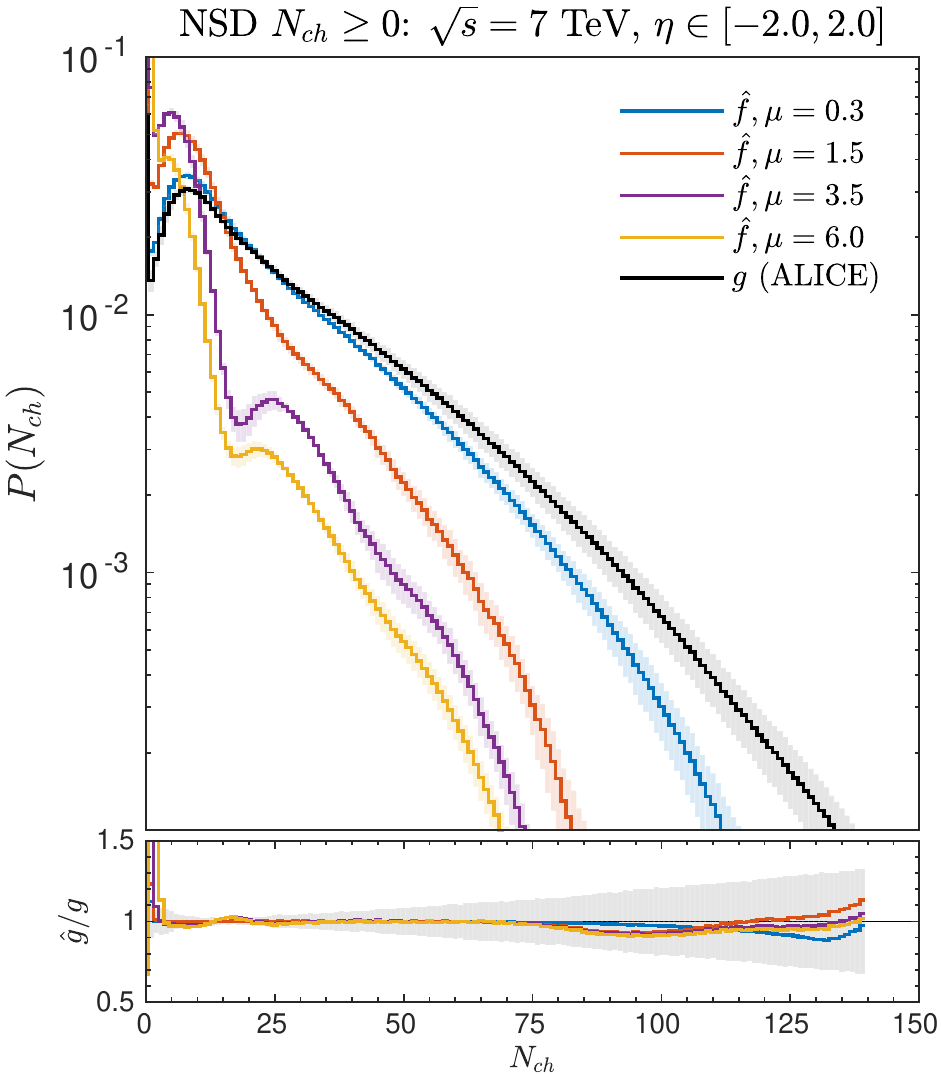} \\
\includegraphics[width=68mm]{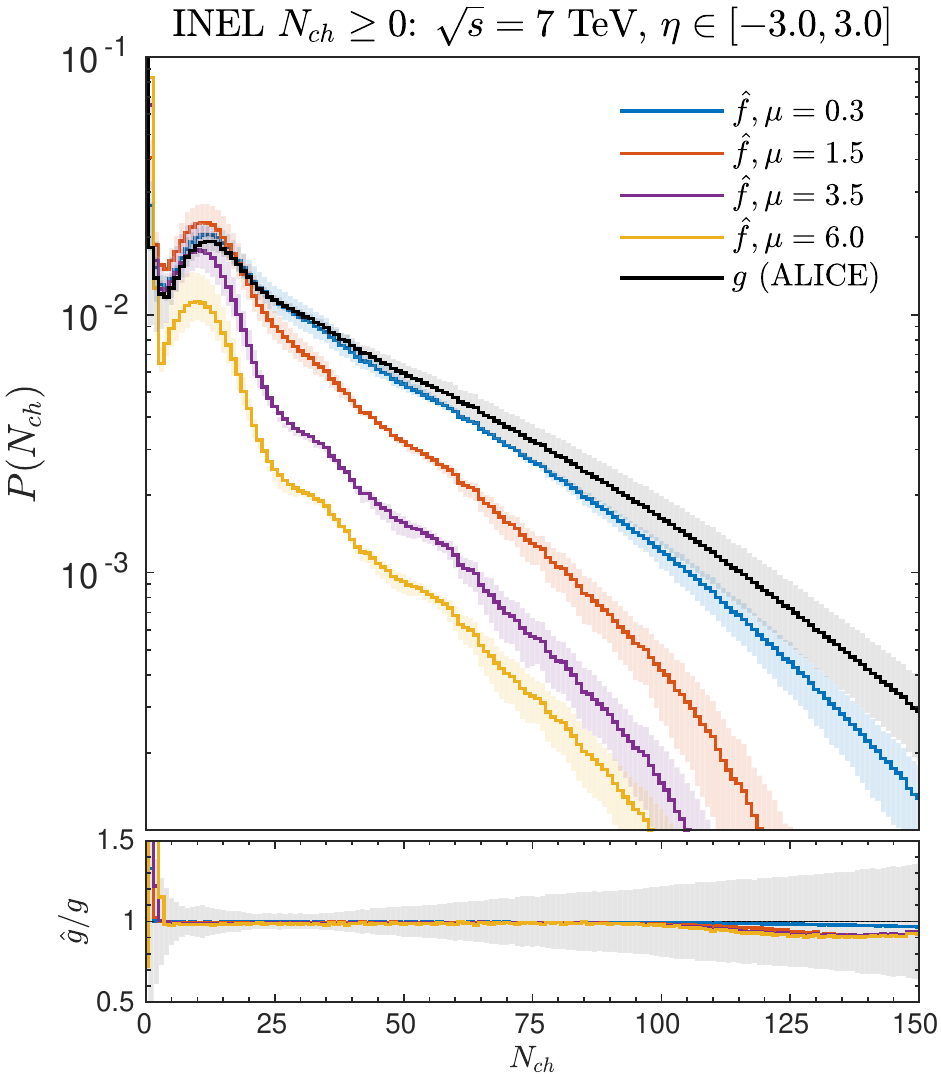} &
\includegraphics[width=68mm]{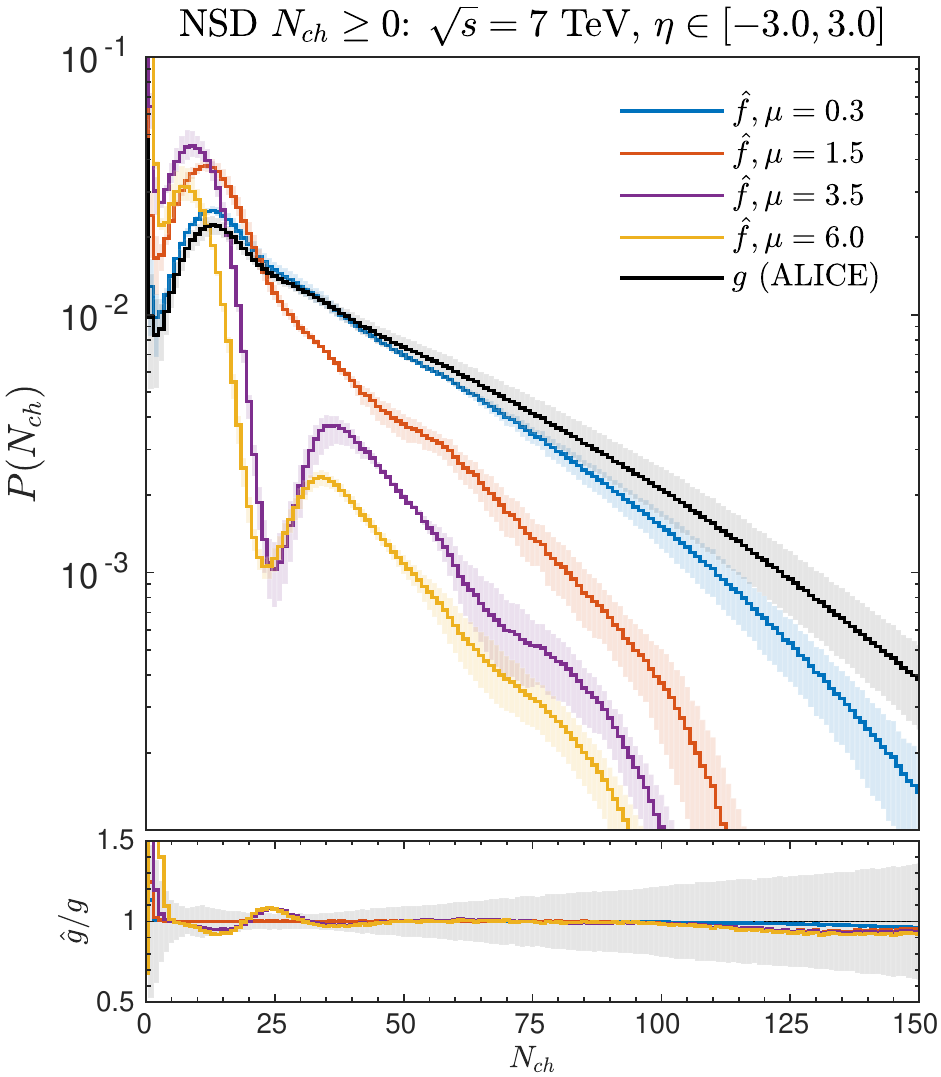} \\
\includegraphics[width=68mm]{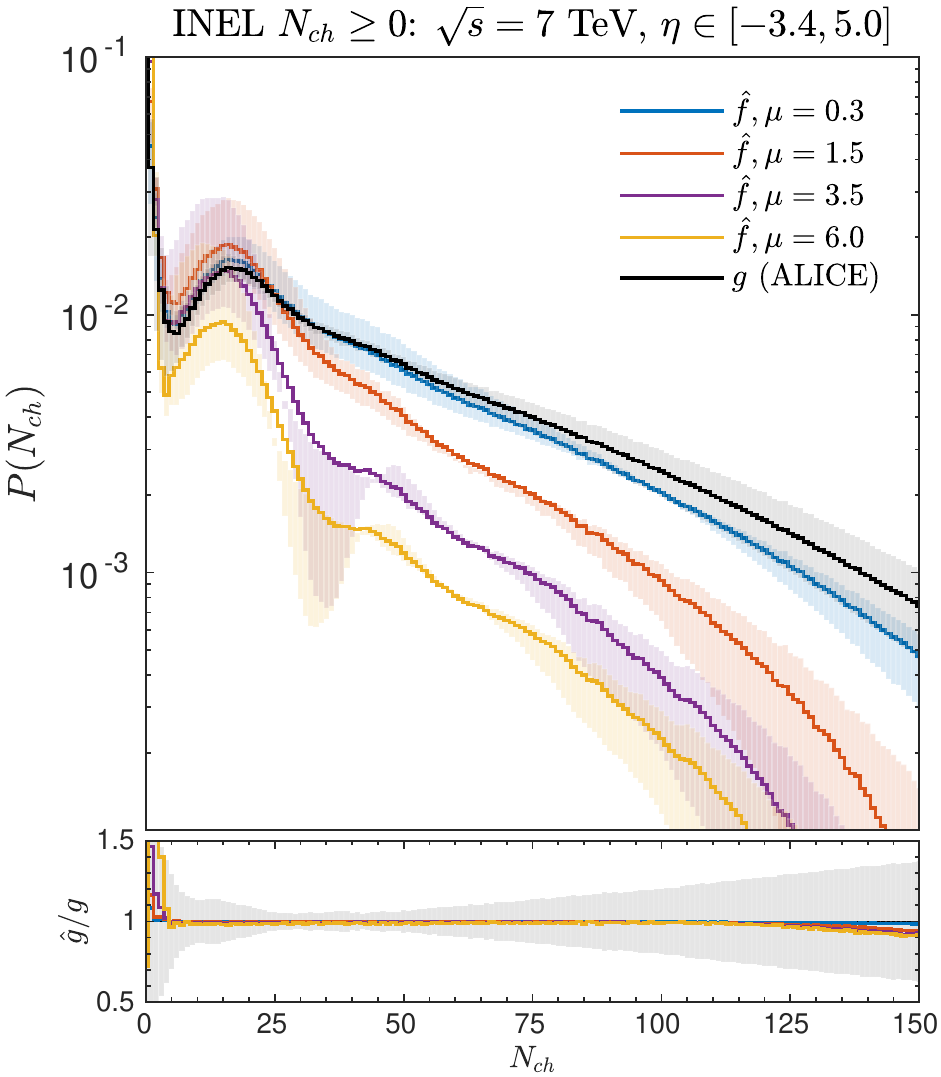} & 
\includegraphics[width=68mm]{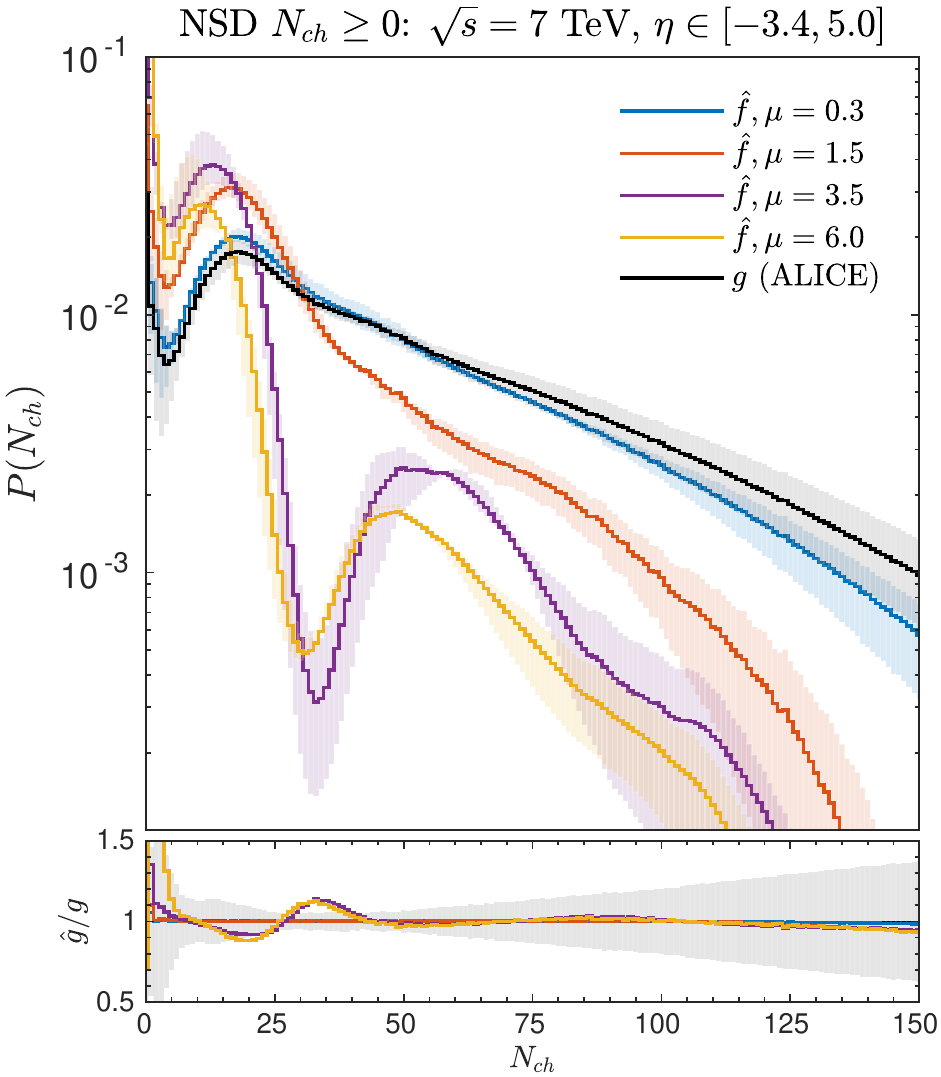} \\
\end{array}$
\end{center}
\caption{ALICE INEL (left) and NSD (right) data at $\sqrt{s} = 7$ TeV \cite{ALICE:2017pcy} for three different pseudorapidity intervals and the \textsc{Kisu} inversion results under different $\mu$-hypothesis.}
\label{fig:ALICE2}
\end{figure}

We executed \textsc{Kisu} inversion simply based on the Poisson compounding hypothesis with different $\mu$ values and the regularization being fixed to small $\lambda = 0.03$ in order not to bias the distribution shapes. Varying regularization strength results in principle in a systematic uncertainty, but we observed it affect mainly the small scale structure. The ALICE data uncertainties are a combination of systematic factors with technically unknown distribution coverage, for example soft QCD Monte Carlo modeling affecting estimated trigger efficiencies and detector unfolding, and statistical counting fluctuations. For the systematic shape variations we did a minimum and maximum global shape shift and evaluated statistical bin-by-bin bootstrap Poisson re-sampling on top of that, with the given event sample sizes $\mathcal{O}(7 \cdot 10^6)$ at 0.9 TeV and $\mathcal{O}(6 \cdot 10^7)$ at 7 TeV from \cite{ALICE:2017pcy}. The uncertainties in Figures \ref{fig:ALICE1} and \ref{fig:ALICE2} represent the 95CL values of this procedure. In principal, one could have done also bin-by-bin bootstrap for the systematic shapes, but because that procedure would neglect all bin-to-bin continuity correlations, it would give way too large high frequency fluctuations not really reflecting typical systematic spectrum distortion or bias variations.

\begin{figure}[H]
\begin{center}
$\begin{array}{ll}
\includegraphics[scale=0.66]{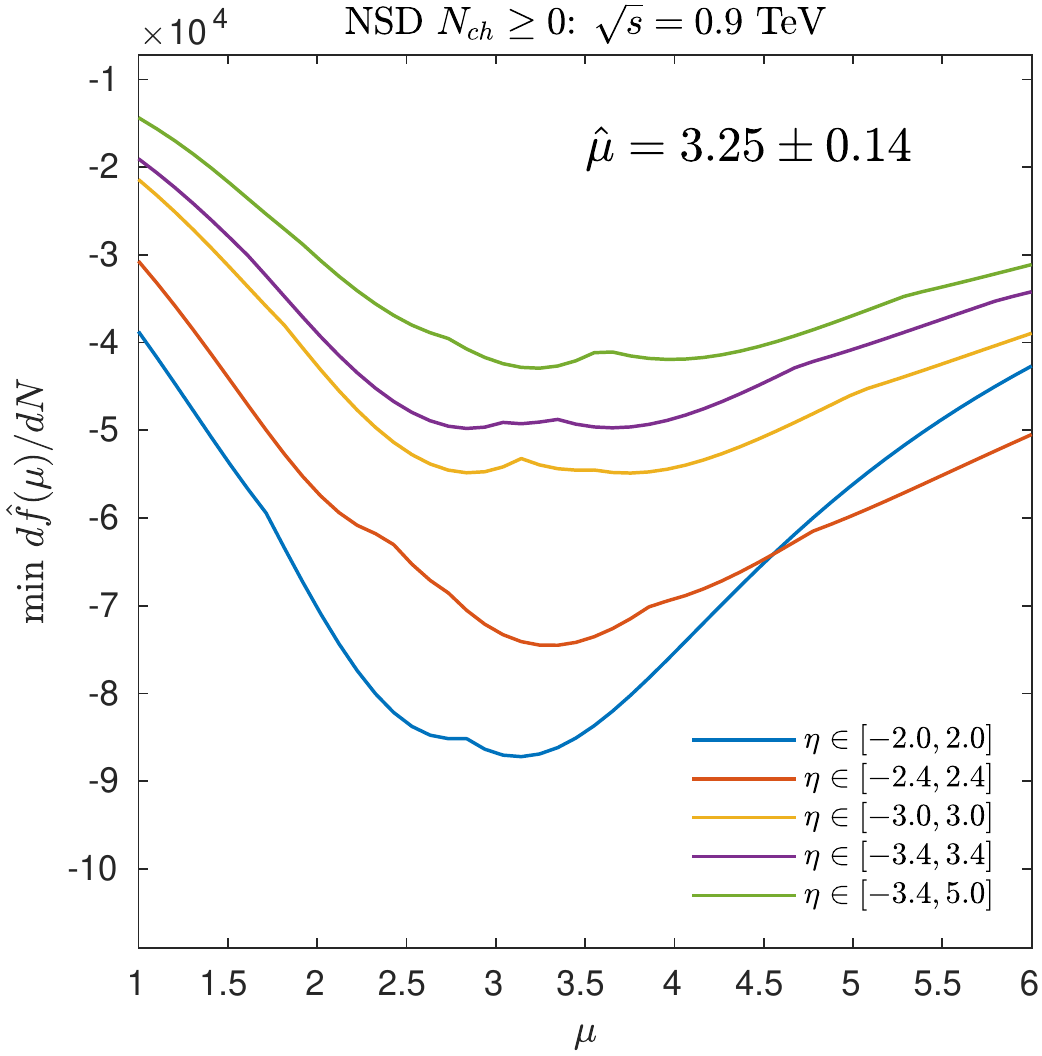} &
\includegraphics[scale=0.66]{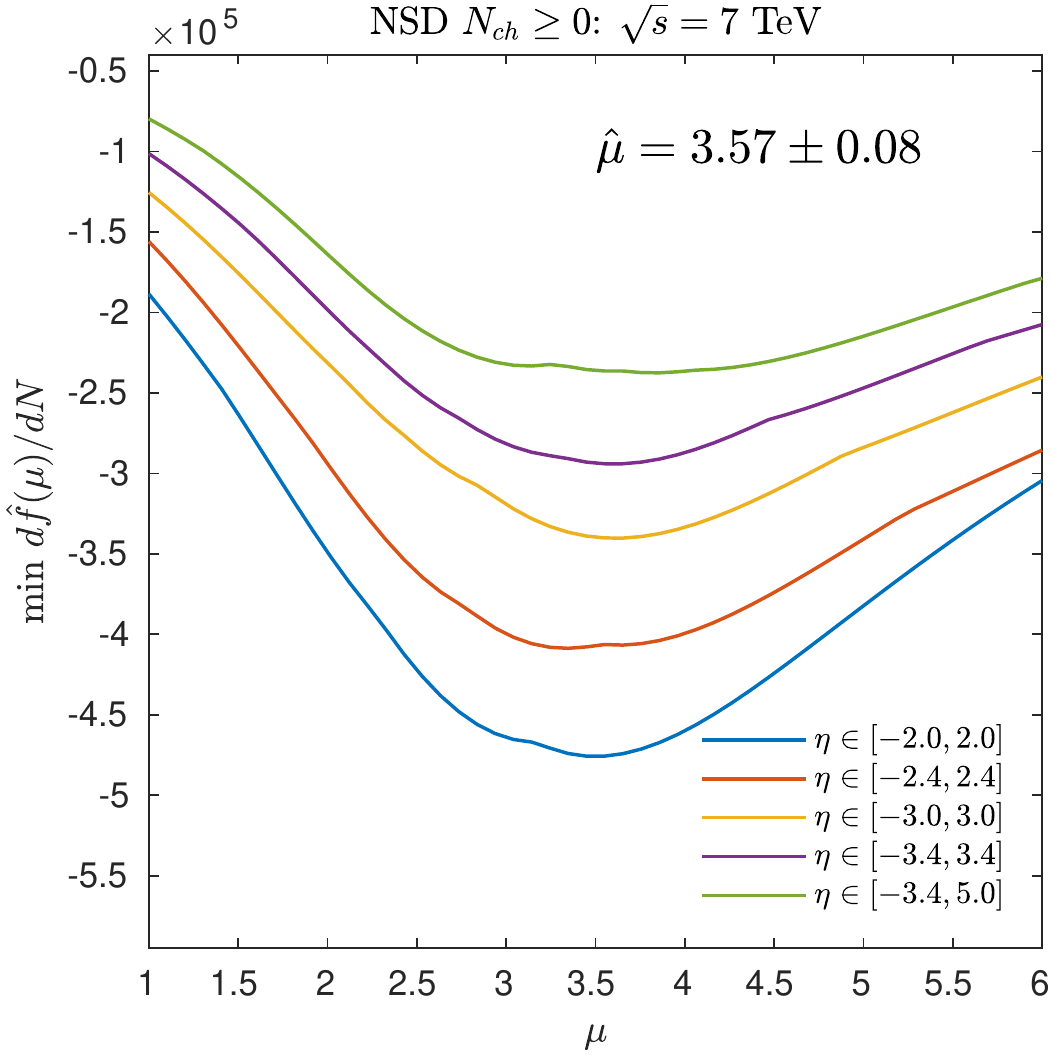}
\end{array}$
\end{center}
\caption{ALICE NSD data parameter $\mu$-scans for two different energies, $\sqrt{s} = 0.9$ TeV (left) and 7 TeV (right).}
\label{fig:muscan}
\end{figure}

As a surprise from the inversion, a strong hidden secondary NBD like peak structure appears at certain $\mu$ hypothesis, most strongly with Poisson $\mu \approx 3 \dots 4$ number of simultaneous `mini-collisions', with the zero-suppressed average given by Equation \ref{eq:muaverage}. Also interesting is that this peak is not strongly visible in the INEL class but appears only once single diffractive events have been suppressed using the NSD class. It seems that the first NBD like peak is similar at both energies, but the second runs with energy. To point out, the INEL with $N_{ch} \geq 1$ event class (figures available using our \href{https://github.com/mieskolainen}{code}) gives results slightly similar to the NSD class, this requirement effectively also suppress diffraction. Also, in general the secondary peak increases in relative amplitude when the pseudorapidity window is enlarged and also the peak position shifts. A thorough interpretation of this observation requires further studies. A typical interpretation of the negative binomial double peak structures is that there are two separate mechanisms for the particle production initiator, soft confining and hard point like. The high multiplicity tail is also interesting, usually difficult to describe perfectly by many Monte Carlo models. For comparisons, see \cite{ALICE:2017pcy}.

In Figure \ref{fig:muscan}, we scan numerically over $\mu$ for the minimum of $\frac{d\hat{f}(\mu)}{dN}$ as the criteria for the steepest dip. The results for $\hat{\mu}$ show the mean and its standard error over different pseudorapidity intervals, value growing with energy, which is expected in the multiparton interaction picture with increasing particle densities. At lower energy, there seems to be larger variation with the solutions. We found out that using larger regularization $\lambda$ values pushes the minimum towards lower values of $\mu$, perhaps simply (over)-smoothing the solutions. Finally, we see in Figure \ref{fig:ALICE1} and \ref{fig:ALICE2} ratio plots that the re-projection $\hat{g} = \mathcal{\hat{F}}(\hat{f})$ versus the measured $g$ has the largest tension at very low multiplicities $N_{ch} < 5$, in the domain dominated by diffraction with qualitatively different behavior than the NBD peaks. The second interesting domain is around $N_{ch} \sim 25$, where oscillations are manifest.

\begin{figure}[H]
\begin{center}
$\begin{array}{ll}
\includegraphics[scale=0.7]{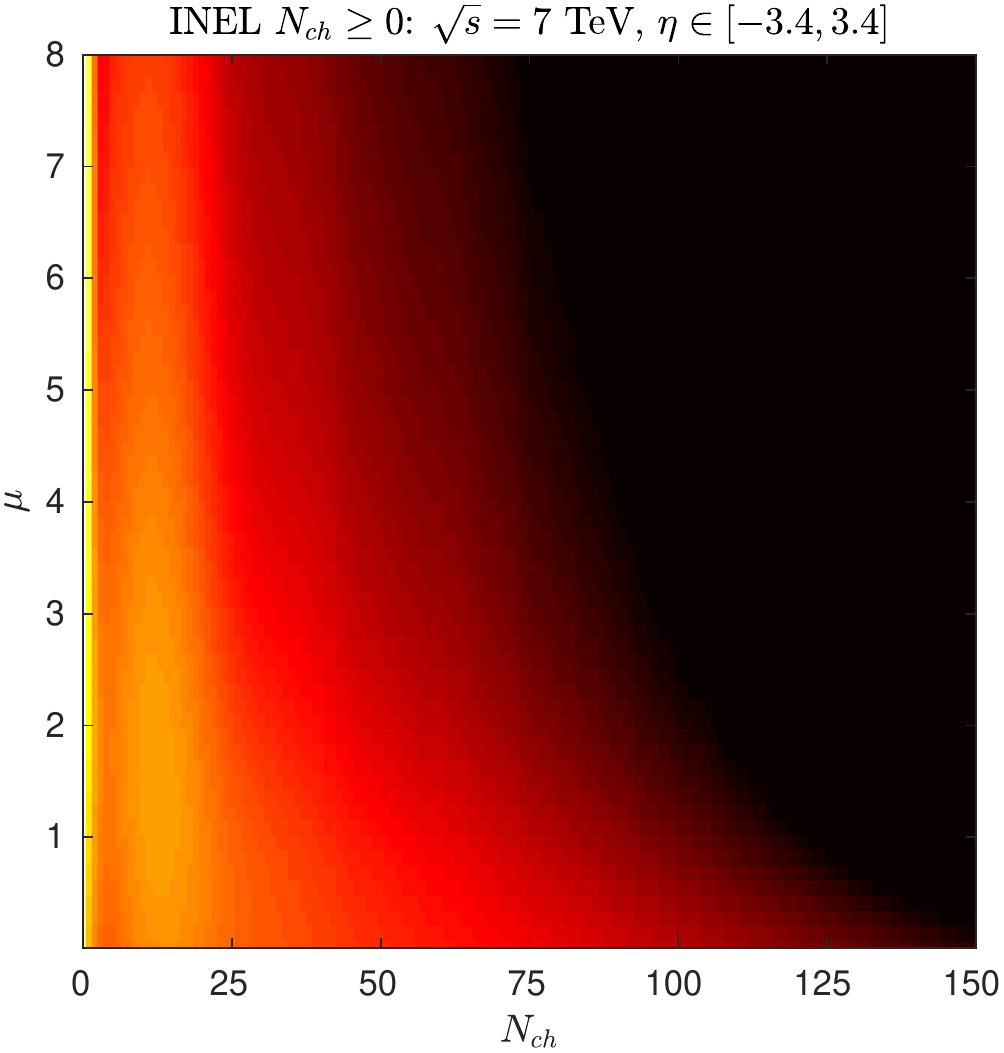} &
\includegraphics[scale=0.7]{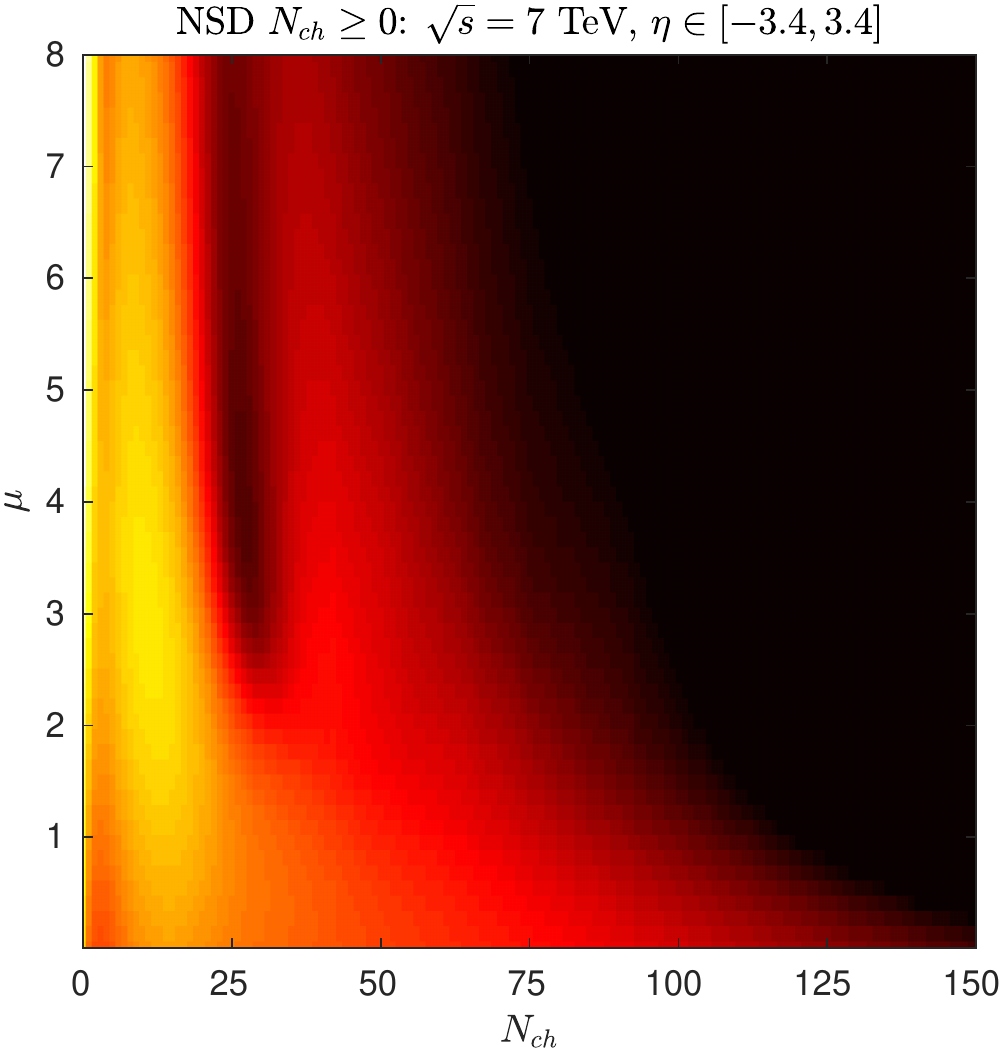}
\end{array}$
\includegraphics[scale=0.7]
{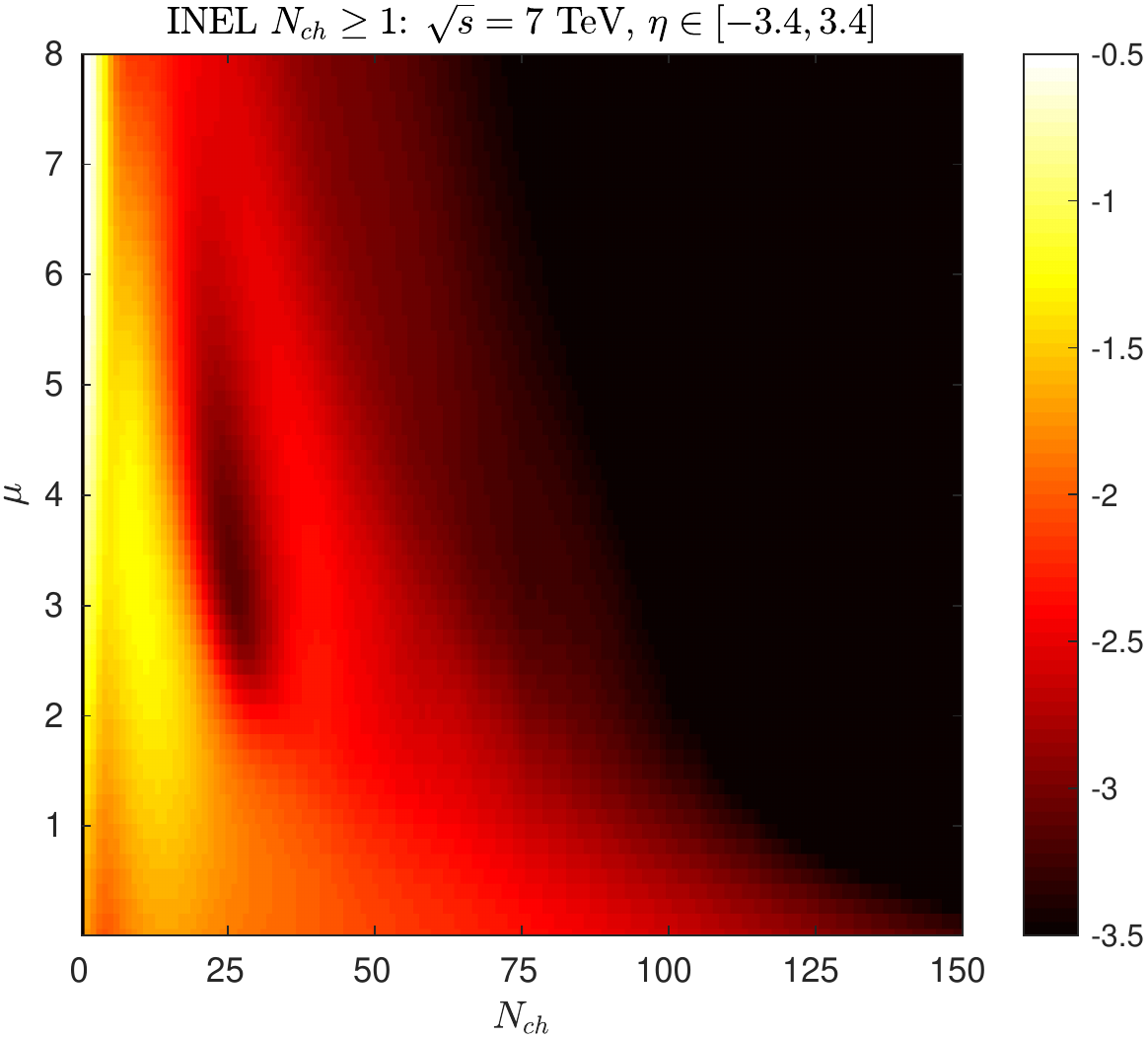}
\end{center}
\caption{Inversion results of $\hat{f}$ in the $(N_{ch},\mu)$ plane at $\sqrt{s} = 7$ TeV for $\eta \in [-3.4,3.4]$. Colors denote values of $\log_{10}[P(N_{ch})]$. }
\label{fig:2dscan}
\end{figure}

Figure \ref{fig:2dscan} demonstrates the `topological' differences between the inversion results as a function of the $\mu$ parameter. The distribution splits and developes the secondary peak at certain $\mu$-value in the case of the NSD $N_{ch} \geq 0$ event class. In the case of the INEL $N_{ch} \geq 1$ splitting also happens, and the re-fusion of two peaks happens faster as a function of the growing $\mu$-value. Where as in the case of the INEL $N_{ch} \geq 0$ class distribution, no splitting is visible.

\section{Conclusions and prospects}
\label{sec:conclusions}

We studied the inversion of a stochastic autoconvolution integral equation and showed that useful inverse solutions can be obtained with a novel algorithm. Special effort was invested in the algorithmic construction from bottom-up and in statistical and systematic uncertainty estimation. A new class of semi-model free phenomenological interpretations and statistical tests of inclusive distributions in soft QCD are possible with the algorithm. Possibly also useful in jet substructure studies when used together with jet algorithms. The algorithm can be also utilized directly as an experimental pileup inversion algorithm in minimum bias studies. There are also possibilities to combine the statistical approach derived here with event-by-event pileup correction algorithms such as \cite{cacciari2015softkiller, bertolini2014pileup}. Future directions may include more detailed analytic treatment in terms of both phenomenology of soft QCD, but also technical development of the algorithm for other type of statistical integral equations. The ultimate goal could be higher dimensional tomography of high energy proton, which may be realistic with the help of deep neural network techniques.

Inverting final state multiplicity spectrum caused by independent multiparton interactions (MPI) is an obvious research target. We did a simple one free parameter inversion using ALICE multiplicity measurements, which exposed an interesting secondary peak structure most prominent with the NSD event selection class. We are not aware of a similar study, typically superposition fits with multiple negative binomial distributions are done, or Monte Carlo model based studies. One could do a further study where different cms-energies and rapidity intervals are matched together using the autoconvolution picture in direct or inverse direction. Speaking in the Regge theory language, approximately similar concept is the multiple independent Pomeron exchanges with `AGK cuts' \cite{abramowsky1974agk}. Given a model of exchange probabilities of multi-Pomeron exchange with inelastic cuts, such as the inelastic eikonal approximation, in principle we can from the measured observables reconstruct a single Pomeron exchange observables without a Monte Carlo model. Deviation from expectations indicate multipomeron (enhanced) interactions which are nonlinear effects not fitting in the $K$-fold autoconvolution picture. Thus, the inverse algorithm result can be seen as a statistical null-hypothesis generator. In addition, probing the fuzzy interface between multiplicities driven by a semi-hard scale interaction versus soft cut Pomerons, could be interesting. Also the nonlinear gluon saturation conjecture at low Bjorken-$x$ and the robust experimental observables of it are open problems.

Combining the approach here with other mathematical tools, the shadow of diffraction can be studied in a new light. That is, it would be interesting to see if data can \textit{directly} show evidence for the AGK cuts like phenomena -- certain specific additive algebraic rules, multiplicity density, rapidity gap topologies. So far, no water-proof observables with one-to-one mapping with AGK have been measured, as far as we know. We also see that methods here could possibly give new insight to the `centrality' determination in heavy ion collisions, when combined together with Gribov-Glauber model. Inverting the measured transverse energy distribution could lead to more precise understanding of pure `binary scaling' (independent nucleon-nucleon collisions) versus strong nuclear effects. The mathematical description here is, essentially, what is approximately meant by binary scaling.

\paragraph{Notes:} The complete code to reproduce all algorithms and figures in this paper is available under the MIT license at \href{https://github.com/mieskolainen}{github.com/mieskolainen}.


\acknowledgments

Risto Orava is acknowledged for discussions on the topic.

\bibliography{mainlatex}

\begin{thebibliography}{48}
\providecommand{\natexlab}[1]{#1}
\providecommand{\url}[1]{\texttt{#1}}
\expandafter\ifx\csname urlstyle\endcsname\relax
  \providecommand{\doi}[1]{doi: #1}\else
  \providecommand{\doi}{doi: \begingroup \urlstyle{rm}\Url}\fi

\bibitem[Abramowsky et~al.(1974)Abramowsky, Gribov, and
  Kancheli]{abramowsky1974agk}
V.~Abramowsky, V.~Gribov, and O.~Kancheli.
\newblock \emph{Sov. J. Nucl. Phys}, 18\penalty0 (308):\penalty0 21, 1974.

\bibitem[Acharya et~al.(2017)]{ALICE:2017pcy}
S.~Acharya et~al.
\newblock {Charged-particle multiplicity distributions over a wide
  pseudorapidity range in proton-proton collisions at $\sqrt{s}=$ 0.9, 7, and 8
  TeV}.
\newblock \emph{Eur. Phys. J.}, C77\penalty0 (12):\penalty0 852, 2017.
\newblock \doi{10.1140/epjc/s10052-017-5412-6}.

\bibitem[Alner et~al.(1985)Alner, Alpg{\aa}rd, Anderer, Ansorge, {\AA}sman,
  B{\"o}ckmann, Booth, Burow, Carlson, Chevalley,
  et~al.]{alner1985multiplicity}
G.~Alner, K.~Alpg{\aa}rd, P.~Anderer, R.~Ansorge, B.~{\AA}sman,
  K.~B{\"o}ckmann, C.~Booth, L.~Burow, P.~Carlson, J.-L. Chevalley, et~al.
\newblock Multiplicity distributions in different pseudorapidity intervals at a
  $\text{CMS}$ energy of 540 $\text{GeV}$.
\newblock \emph{Physics Letters B}, 160\penalty0 (1-3):\penalty0 193--198,
  1985.

\bibitem[Andersson et~al.(1979)Andersson, Gustafson, and
  Peterson]{andersson1979semiclassical}
B.~Andersson, G.~Gustafson, and C.~Peterson.
\newblock A semiclassical model for quark jet fragmentation.
\newblock \emph{Zeitschrift f{\"u}r Physik C Particles and Fields}, 1\penalty0
  (1):\penalty0 105--116, 1979.

\bibitem[Andersson et~al.(1983)Andersson, Gustafson, Ingelman, and
  Sj{\"o}strand]{andersson1983parton}
B.~Andersson, G.~Gustafson, G.~Ingelman, and T.~Sj{\"o}strand.
\newblock Parton fragmentation and string dynamics.
\newblock \emph{Physics Reports}, 97\penalty0 (2-3):\penalty0 31--145, 1983.

\bibitem[Andersson et~al.(1989)Andersson, Dahlqvist, and
  Gustafson]{andersson1989local}
B.~Andersson, P.~Dahlqvist, and G.~Gustafson.
\newblock On local parton-hadron duality.
\newblock \emph{Zeitschrift f{\"u}r Physik C Particles and Fields}, 44\penalty0
  (3):\penalty0 461--466, 1989.

\bibitem[Artru and Mennessier(1974)]{artru1974string}
X.~Artru and G.~Mennessier.
\newblock String model and multiproduction.
\newblock \emph{Nuclear Physics B}, 70\penalty0 (1):\penalty0 93--115, 1974.

\bibitem[Azimov et~al.(1985)Azimov, Dokshitzer, Khoze, and
  Trovan]{azimov1985similarity}
Y.~I. Azimov, Y.~L. Dokshitzer, V.~A. Khoze, and S.~Trovan.
\newblock Similarity of parton and hadron spectra in $\text{QCD}$ jets.
\newblock \emph{Zeitschrift f{\"u}r Physik C Particles and Fields}, 27\penalty0
  (1):\penalty0 65--72, 1985.

\bibitem[Bardsley and Vogel(2004)]{bardsley2004nonnegatively}
J.~M. Bardsley and C.~R. Vogel.
\newblock A nonnegatively constrained convex programming method for image
  reconstruction.
\newblock \emph{SIAM Journal on Scientific Computing}, 25\penalty0
  (4):\penalty0 1326--1343, 2004.

\bibitem[Bertolini et~al.(2014)Bertolini, Harris, Low, and
  Tran]{bertolini2014pileup}
D.~Bertolini, P.~Harris, M.~Low, and N.~Tran.
\newblock Pileup per particle identification.
\newblock \emph{Journal of High Energy Physics}, 2014\penalty0 (10):\penalty0
  59, 2014.

\bibitem[Blobel(2002)]{blobel2002unfolding}
V.~Blobel.
\newblock An unfolding method for high energy physics experiments.
\newblock \emph{arXiv preprint hep-ex/0208022}, 2002.

\bibitem[B{\o}gsted and Pitts(2010)]{bogsted2010decompounding}
M.~B{\o}gsted and S.~M. Pitts.
\newblock Decompounding random sums: a nonparametric approach.
\newblock \emph{Annals of the Institute of Statistical Mathematics},
  62\penalty0 (5):\penalty0 855--872, 2010.

\bibitem[Cacciari et~al.(2015)Cacciari, Salam, and
  Soyez]{cacciari2015softkiller}
M.~Cacciari, G.~P. Salam, and G.~Soyez.
\newblock $\text{SoftKiller}$, a particle-level pileup removal method.
\newblock \emph{The European Physical Journal C}, 75\penalty0 (2):\penalty0 59,
  2015.

\bibitem[Choi and Lanterman(2005)]{choi2005iterative}
K.~Choi and A.~D. Lanterman.
\newblock An iterative deautoconvolution algorithm for nonnegative functions.
\newblock \emph{Inverse problems}, 21\penalty0 (3):\penalty0 981, 2005.

\bibitem[Csiszar(1991)]{csiszar1991least}
I.~Csiszar.
\newblock Why least squares and maximum entropy? $\text{An}$ axiomatic approach
  to inference for linear inverse problems.
\newblock \emph{The Annals of Statistics}, pages 2032--2066, 1991.

\bibitem[D'Agostini(1995)]{d1995multidimensional}
G.~D'Agostini.
\newblock A multidimensional unfolding method based on $\text{Bayes'}$ theorem.
\newblock \emph{Nuclear Instruments and Methods in Physics Research Section A:
  Accelerators, Spectrometers, Detectors and Associated Equipment},
  362\penalty0 (2):\penalty0 487--498, 1995.

\bibitem[Dempster et~al.(1977)Dempster, Laird, and Rubin]{dempster1977maximum}
A.~P. Dempster, N.~M. Laird, and D.~B. Rubin.
\newblock Maximum likelihood from incomplete data via the $\text{EM}$
  algorithm.
\newblock \emph{Journal of the royal statistical society. Series B
  (methodological)}, pages 1--38, 1977.

\bibitem[Dokshitzer et~al.(1991)Dokshitzer, Khoze, Mueller, and
  Troyan]{dokshitzer1991basics}
Y.~Dokshitzer, V.~Khoze, A.~Mueller, and S.~Troyan.
\newblock \emph{Basics of perturbative QCD}.
\newblock Atlantica S{\'e}guier Fronti{\`e}res, 1991.

\bibitem[Efron(1979)]{efron1979bootstrap}
B.~Efron.
\newblock Bootstrap methods:$\{$A$\}$nother look at the jackknife.
\newblock \emph{The Annals of Statistics}, 7:\penalty0 1--26, 1979.

\bibitem[Efron and Tibshirani(1994)]{efron1994introduction}
B.~Efron and R.~J. Tibshirani.
\newblock \emph{An Introduction to the Bootstrap}.
\newblock CRC press, 1994.

\bibitem[Feynman(1969)]{feynman1969very}
R.~P. Feynman.
\newblock Very high-energy collisions of hadrons.
\newblock \emph{Physical Review Letters}, 23\penalty0 (24):\penalty0 1415,
  1969.

\bibitem[Field and Feynman(1978)]{field1978parametrization}
R.~D. Field and R.~P. Feynman.
\newblock A parametrization of the properties of quark jets.
\newblock \emph{Nuclear Physics B}, 136\penalty0 (1):\penalty0 1--76, 1978.

\bibitem[Gerth et~al.(2014)Gerth, Hofmann, Birkholz, Koke, and
  Steinmeyer]{gerth2014regularization}
D.~Gerth, B.~Hofmann, S.~Birkholz, S.~Koke, and G.~Steinmeyer.
\newblock Regularization of an autoconvolution problem in ultrashort laser
  pulse characterization.
\newblock \emph{Inverse Problems in Science and Engineering}, 22\penalty0
  (2):\penalty0 245--266, 2014.

\bibitem[Ghosh(2012)]{ghosh2012negative}
P.~Ghosh.
\newblock Negative binomial multiplicity distribution in proton-proton
  collisions in limited pseudorapidity intervals at $\text{LHC}$ up to
  $\sqrt{s} = 7$ $\text{TeV}$ and the clan model.
\newblock \emph{Physical Review D}, 85\penalty0 (5):\penalty0 054017, 2012.

\bibitem[Giovannini and Van~Hove(1986)]{giovannini1986negative}
A.~Giovannini and L.~Van~Hove.
\newblock Negative binomial multiplicity distributions in high energy hadron
  collisions.
\newblock \emph{Zeitschrift f{\"u}r Physik C Particles and Fields}, 30\penalty0
  (3):\penalty0 391--400, 1986.

\bibitem[Gorenflo and Hofmann(1994)]{gorenflo1994autoconvolution}
R.~Gorenflo and B.~Hofmann.
\newblock On autoconvolution and regularization.
\newblock \emph{Inverse Problems}, 10\penalty0 (2):\penalty0 353, 1994.

\bibitem[Grubbstr{\"o}m and Tang(2006)]{grubbstrom2006moments}
R.~W. Grubbstr{\"o}m and O.~Tang.
\newblock The moments and central moments of a compound distribution.
\newblock \emph{European Journal of Operational Research}, 170\penalty0
  (1):\penalty0 106--119, 2006.

\bibitem[Hagedorn(1965)]{hagedorn1965statistical}
R.~Hagedorn.
\newblock Statistical thermodynamics of strong interactions at high energies.
\newblock \emph{Nuovo Cimento, Suppl.}, 3\penalty0 (CERN-TH-520):\penalty0
  147--186, 1965.

\bibitem[Hall(1986)]{hall1986bootstrap}
P.~Hall.
\newblock On the bootstrap and confidence intervals.
\newblock \emph{The Annals of Statistics}, pages 1431--1452, 1986.

\bibitem[Hall(2013)]{hall2013bootstrap}
P.~Hall.
\newblock \emph{The bootstrap and $\text{Edgeworth}$ expansion}.
\newblock Springer Science \& Business Media, 2013.

\bibitem[Kaidalov(1982)]{kaidalov1982quark}
A.~Kaidalov.
\newblock The quark-gluon structure of the pomeron and the rise of inclusive
  spectra at high energies.
\newblock \emph{Physics Letters B}, 116\penalty0 (6):\penalty0 459--463, 1982.

\bibitem[Kittel and De~Wolf(2005)]{kittel2005soft}
W.~Kittel and E.~A. De~Wolf.
\newblock \emph{Soft multihadron dynamics}.
\newblock World Scientific, 2005.

\bibitem[Koba et~al.(1972)Koba, Nielsen, and Olesen]{koba1972scaling}
Z.~Koba, H.~B. Nielsen, and P.~Olesen.
\newblock Scaling of multiplicity distributions in high energy hadron
  collisions.
\newblock \emph{Nuclear Physics B}, 40:\penalty0 317--334, 1972.

\bibitem[Kullback and Leibler(1951)]{kullback1951information}
S.~Kullback and R.~A. Leibler.
\newblock On information and sufficiency.
\newblock \emph{The Annals of Mathematical Statistics}, 22\penalty0
  (1):\penalty0 79--86, 1951.

\bibitem[Kuusela and Panaretos(2015)]{kuusela2015statistical}
M.~Kuusela and V.~M. Panaretos.
\newblock Statistical unfolding of elementary particle spectra:
  $\text{Empirical Bayes}$ estimation and bias-corrected uncertainty
  quantification.
\newblock \emph{The Annals of Applied Statistics}, 9\penalty0 (3):\penalty0
  1671--1705, 2015.

\bibitem[Landweber(1951)]{landweber1951iteration}
L.~Landweber.
\newblock An iteration formula for $\text{Fredholm}$ integral equations of the
  first kind.
\newblock \emph{American journal of mathematics}, 73\penalty0 (3):\penalty0
  615--624, 1951.

\bibitem[Lucy(1974)]{lucy1974iterative}
L.~B. Lucy.
\newblock An iterative technique for the rectification of observed
  distributions.
\newblock \emph{The astronomical journal}, 79:\penalty0 745, 1974.

\bibitem[Martin et~al.(2009)Martin, O'Bryant, et~al.]{martin2009supremum}
G.~Martin, K.~O'Bryant, et~al.
\newblock The supremum of autoconvolutions, with applications to additive
  number theory.
\newblock \emph{Illinois Journal of Mathematics}, 53\penalty0 (1):\penalty0
  219--235, 2009.

\bibitem[Matthews(2011)]{matthews2011thesis}
Z.~L. Matthews.
\newblock Proton-proton collisions at the $\text{Large Hadron Collider's}$
  $\text{ALICE}$ $\text{E}$xperiment: diffraction and high multiplicity,
  $\text{P}$h$\text{D}$ thesis.
\newblock December 2011.

\bibitem[Meister(2007)]{meister2007optimal}
A.~Meister.
\newblock Optimal convergence rates for density estimation from grouped data.
\newblock \emph{Statistics \& probability letters}, 77\penalty0 (11):\penalty0
  1091--1097, 2007.

\bibitem[Moskow and Schotland(2009)]{moskow2009numerical}
S.~Moskow and J.~C. Schotland.
\newblock Numerical studies of the inverse $\text{Born}$ series for diffuse
  waves.
\newblock \emph{Inverse Problems}, 25\penalty0 (9):\penalty0 095007, 2009.

\bibitem[Mueller and Siltanen(2012)]{mueller2012linear}
J.~L. Mueller and S.~Siltanen.
\newblock \emph{Linear and nonlinear inverse problems with practical
  applications}, volume~10.
\newblock Siam, 2012.

\bibitem[Polyakov(1970)]{polyakov1970hypothesis}
A.~Polyakov.
\newblock Hypothesis of self-similarity in strong interactions: 1.
  $\text{P}$lural generation of e+e- annihilation hadrons.
\newblock \emph{Zh. Eksp. Teor. Fiz}, 59:\penalty0 542, 1970.

\bibitem[Richardson(1972)]{richardson1972bayesian}
W.~H. Richardson.
\newblock Bayesian-based iterative method of image restoration.
\newblock \emph{JOSA}, 62\penalty0 (1):\penalty0 55--59, 1972.

\bibitem[Van~Es et~al.(2007)Van~Es, Gugushvili, Spreij, et~al.]{van2007kernel}
B.~Van~Es, S.~Gugushvili, P.~Spreij, et~al.
\newblock A kernel type nonparametric density estimator for decompounding.
\newblock \emph{Bernoulli}, 13\penalty0 (3):\penalty0 672--694, 2007.

\bibitem[Webber(1984)]{webber1984qcd}
B.~R. Webber.
\newblock A $\text{QCD}$ model for jet fragmentation including soft gluon
  interference.
\newblock \emph{Nuclear Physics B}, 238\penalty0 (3):\penalty0 492--528, 1984.

\bibitem[Wiener(1949)]{wiener1949extrapolation}
N.~Wiener.
\newblock \emph{Extrapolation, interpolation, and smoothing of stationary time
  series}, volume~7.
\newblock MIT press Cambridge, MA, 1949.

\bibitem[Zech(2016)]{zech2016analysis}
G.~Zech.
\newblock Analysis of distorted measurements--parameter estimation and
  unfolding.
\newblock \emph{arXiv preprint arXiv:1607.06910}, 2016.

\end{thebibliography}




\end{document}